\documentclass[a4paper,11pt]{article}
\pdfoutput=1 
             
\usepackage{jcappub}
\usepackage{ulem}

\usepackage{tcolorbox}

\title{QUBIC II: Spectral Polarimetry with Bolometric Interferometry}

\author[1]{L.~Mousset}
\author[2,3]{M.M.~Gamboa Lerena}
\author[4,5]{E.S.~Battistelli}
\author[4,5]{P.~de~Bernardis}
\author[1]{P.~Chanial}
\author[4,5]{G.~D'Alessandro}
\author[6]{G.~Dashyan}
\author[4,5]{M.~De~Petris}
\author[1]{L.~Grandsire}
\author[1]{J.-Ch.~Hamilton}
% ...

\affiliation[1]{Universit\'e de Paris, CNRS, Astroparticule et Cosmologie, F-75006 Paris, France}
\affiliation[2]{Facultad de Ciencias Astron\a'{o}micas y Geof\a'{i}sicas (Universidad Nacional de La Plata), Argentina}
\affiliation[3]{CONICET, Argentina}
\affiliation[4]{Universit\a`{a} di Roma - La Sapienza, Roma, Italy}
\affiliation[5]{INFN sezione di Roma, 00185 Roma, Italy}
\affiliation[6]{Institut d'Astrophysique Spatiale, Orsay (CNRS-INSU), France}
% ...

\emailAdd{mousset@apc.in2p3.fr}
\emailAdd{mgamboa@fcaglp.unlp.edu.ar}

\abstract{

Bolometric interferometry is a novel technique that has the ability to perform spectral imaging. A bolometric interferometer observes the sky in a wide frequency band and can reconstruct sky maps in several sub-bands within the physical band in post-processing of the data. This provides a powerful spectral method to discriminate between the cosmic microwave background (CMB) and astrophysical foregrounds. In this paper, the methodology is illustrated with examples based on the Q \& U Bolometric Interferometer for Cosmology (QUBIC) which is a ground-based instrument designed to measure the B-mode polarization of the sky at millimeter wavelengths. We consider the specific cases of point source reconstruction and Galactic dust mapping and we characterize the point spread function as a function of frequency. We study the noise properties of spectral imaging, especially the correlations between sub-bands, using end-to-end simulations together with a fast noise simulator. We conclude showing that spectral imaging performance are nearly optimal up to five sub-bands in the case of QUBIC.
}


\arxivnumber{2010.15119}

\begin{document} 

\maketitle
\date{Received November 24, 2020; Accepted July 6, 2021}
   
\section{Introduction}

This article is the second in a series of eight on the Q \& U Bolometric Interferometer for Cosmology (QUBIC), and it introduces the spectroscopic imaging capability made possible by bolometric interferometry. QUBIC will observe the sky at millimeter wavelengths, looking at the cosmic microwave background (CMB).

The CMB, detected by Penzias and Wilson~\cite{penzias1965measurement}, has a thermal distribution with a temperature of $2.7255 \pm 0.0006$~K~\cite{fixsen2009temperature}. It is a partially polarized photon field released 380,000 years after the Big Bang when neutral hydrogen was formed. The polarization can be fully described by a scalar and a pseudo-scalar fields: E and B-modes respectively. B-mode polarization anisotropies are generated by primordial gravitational waves occurring at the inflation era.  The indirect detection of these waves would represent a major step towards understanding the inflationary epoch that is believed to have occurred in the early Universe. Tensor modes in the metric perturbations are a specific prediction of inflation. The measurement of the corresponding B-mode polarization anisotropies would reveal the inflationary energy scale, which is directly related to the amplitude of this signal. This amplitude, relative to the scalar mode, is parametrized by the so called tensor-to-scalar ratio $r$.

Currently, there are several instruments aiming at measuring the primordial B modes. These include SPTPol~\cite{2020PhRvD.101l2003S}, SPT-3G~\cite{2021arXiv210306166G}, POLARBEAR~\cite{2020ApJ...897...55P}, ACTpol~\cite{2020JCAP...12..045C,2020JCAP...12..047A}, CLASS~\cite{2020ApJ...891..134X}, 
and BICEP2/Keck Array~\cite{2018PhRvL.121v1301B}. Some of them are planned to be upgraded: CLASS~\cite{2020JLTP..199..289D}, POLARBEAR 2 + Simons Array~\cite{2016JLTP..184..805S}, AdvACT~\cite{AdvACT2016}, and BICEP3/BICEP array~\cite{2020JLTP..199..976S}. Planned experiments include Simons Observatory~\cite{2019JCAP...02..056A}, PIPER~\cite{2021RScI...92c5111D}, LSPE~\cite{2020arXiv200811049T}, CMB-S4~\cite{2019arXiv190704473A} and LiteBIRD~\cite{2020JLTP..199.1107S}. 

The main foreground contaminant at high frequencies is the polarized thermal emission from elongated dust grains in the Galaxy~\cite{akrami2020planck}. At lower frequencies, emission from synchrotron~\cite{fuskeland2014spatial} is expected to be significant, even in a so-called clean CMB field, if $r$ is below $10^{-2}$. Current estimates depend strongly on the assumption that synchrotron is well described by a simple power law with a steep spectral index (but spectral curvature might well be there). While free-free is not expected to be a major contaminant due to its vanishingly small degree of polarization~\cite{macellari2011galactic}, spinning dust, whose impact is addressed by Remazeilles \textit{et al.}~\cite{remazeilles2016sensitivity}, should not be neglected. Since 2014, large advances in understanding the galactic foregrounds and developing mitigation techniques have been made but a lot of work remains to be done in this field and it is a among the main challenges for CMB experiments today.

The control of contamination from foregrounds can only be achieved with a number of frequencies around the maximum emission of the CMB relying on the fact that the spectral distribution of the CMB polarization is significantly different from that of the foregrounds so that low and high frequencies can be used as templates to remove the contamination in the CMB frequencies.

A bolometric interferometer such as QUBIC acquires data in a focal plane containing bolometers operating over a single wide frequency band. However, this type of instrument has the ability to split, in post-processing of the data, the wide band into multiple frequency sub-bands, achieving spectral resolution. In the following, this technique will be called spectral imaging. This idea was already mentioned in earlier work by Malu \textit{et al.}~\cite{malu2010broadband}. It is possible thanks to the very particular instrument beam pattern, given by the geometric distribution of an array of apertures operating as pupils of the interferometer. The synthesized beam contains multiple peaks whose angular separation is linearly dependent on the wavelength.

This paper does not attempt to make realistic simulations and forecasts for the QUBIC project, this is addressed in the companion paper Hamilton \textit{et al.}~\cite{2020.QUBIC.PAPER1}. The goal is to demonstrate spectral imaging technique on simple cases, with very basic foreground models, not accounting for any systematic effect. The expected spectral imaging performance for a real instrument will be treated in detail in the near future.

This article is organized as follows. In section~\ref{BIsection} we describe the working principle of a bolometric interferometer taking the example and characteristics of QUBIC. In section~\ref{sec:specSection} we describe the spectral dependence of the synthesized beam and show under which conditions the spectral information can be recovered at the map making level. In section~\ref{sec:testSection} we test spectral imaging on simple cases using the QUBIC data analysis and simulation pipeline. Finally, in section~\ref{sec:noiseSection} we present the performance of the spectral reconstruction.
We compare a simulated sky to the reconstructed one using the QUBIC data analysis pipeline. Tests with a real point source were carried out in the lab and results are presented by Torchinsky \textit{et al.}~\cite{2020.QUBIC.PAPER3}. 

Detailed information about QUBIC can be found in the companion papers: scientific overview and expected performance of QUBIC~\cite{2020.QUBIC.PAPER1}, characterization of the technological demonstrator (TD)~\cite{2020.QUBIC.PAPER3}, transition-edge sensors and readout characterization~\cite{2020.QUBIC.PAPER4}, cryogenic system performance~\cite{2020.QUBIC.PAPER5}, half-wave plate (HWP) rotator design and performance~\cite{2020.QUBIC.PAPER6}, feed-horn-switches system of the TD~\cite{2020.QUBIC.PAPER7}, and optical design and performance~\cite{2020.QUBIC.PAPER8}.

\section{Bolometric interferometry as synthesized imaging}\label{BIsection}

\begin{figure}[ht!]
\centering
\includegraphics[width=0.65\hsize]{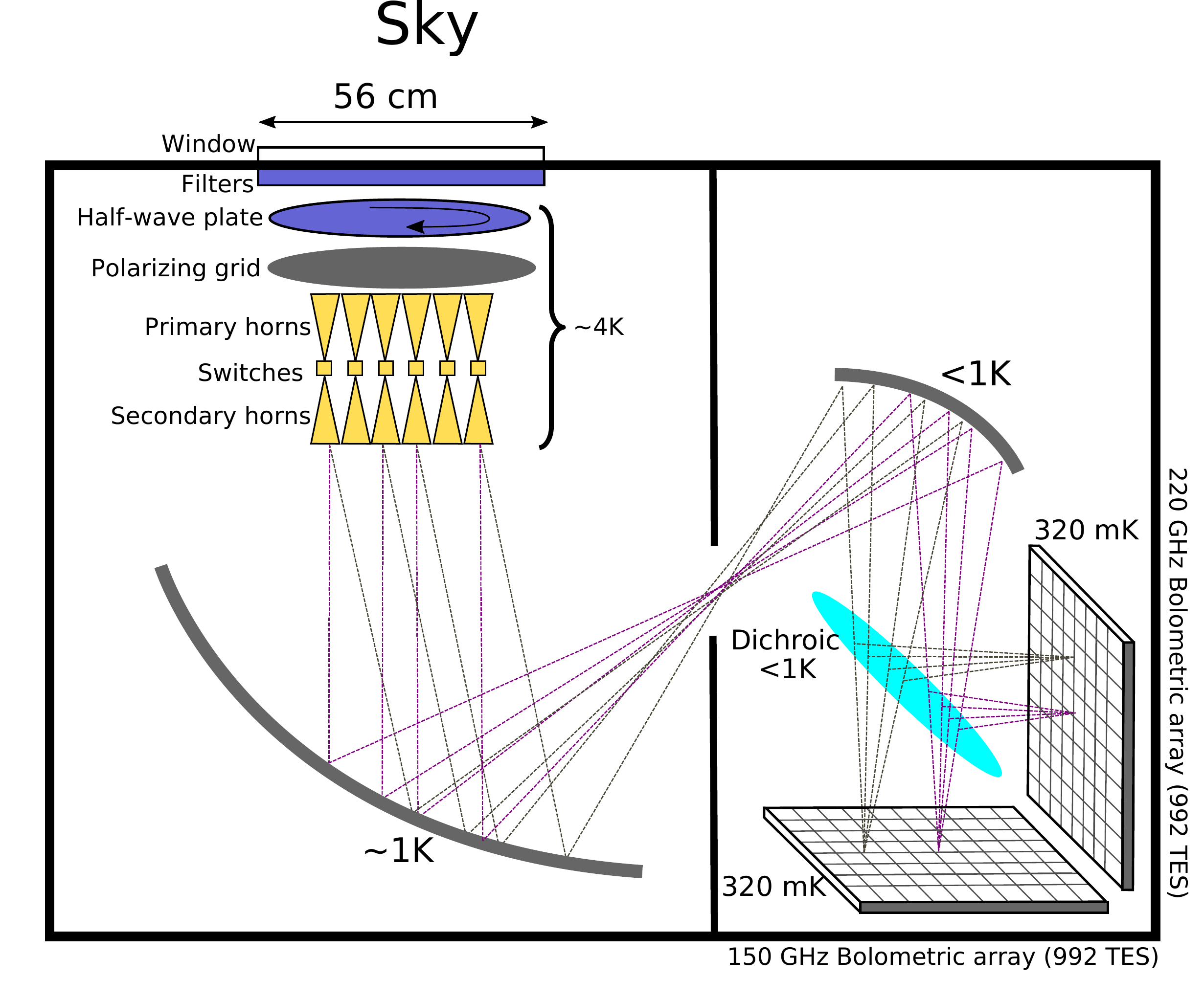}
\includegraphics[width=0.65\hsize]{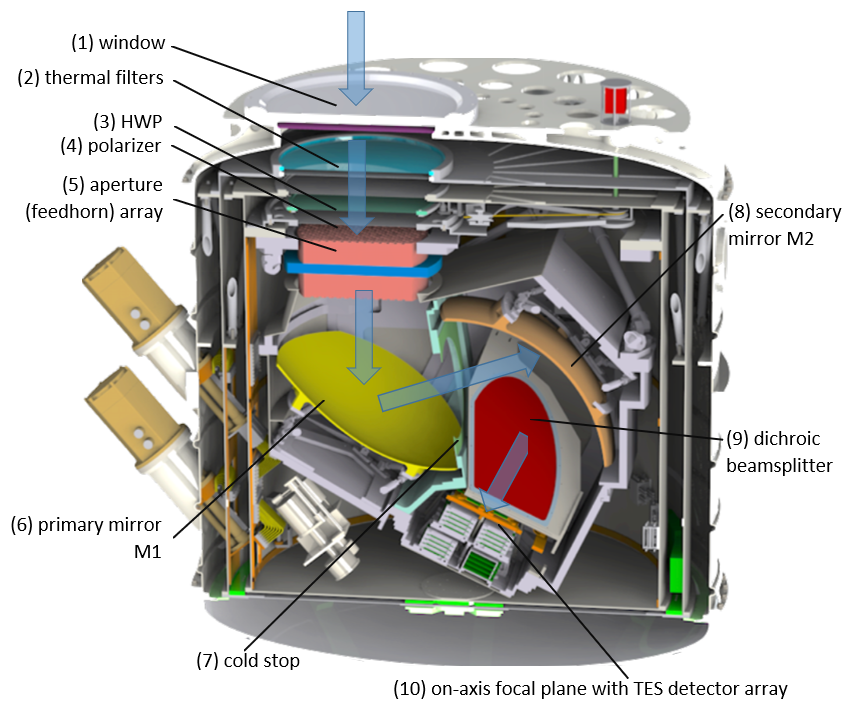}
   \caption{Schematic of the QUBIC instrument and sectional cut of the cryostat showing the same sub-systems in their real configuration.}
   \label{fig:QUBIC_sketch}
\end{figure}

A bolometric interferometer is an instrument observing in the millimeter and submillimeter frequency range based on the Fizeau interferometer~\cite{1851fizeau}. A set of pupils are used at the front of the instrument to select baselines. The resulting interference pattern is then imaged on a focal plane populated with an array of bolometers. With the addition of a polarizing grid and a rotating HWP before the pupil array, the instrument becomes a polarimeter, such as QUBIC. A schematic of the QUBIC instrument optical chain is shown in figure~\ref{fig:QUBIC_sketch} (left) with a sectional cut of the cryostat (right). The main parameters are summarized in table~\ref{tab:qubic_params}.

\begin{table}[t!]
    \renewcommand{\arraystretch}{1.}
    \begin{center}
        \begin{tabular}{p{5cm}  p{5.5cm}}
            \hline
            Parameter & Full instrument value \\
            \hline
            \hline
            Frequency channels \dotfill  & 150 GHz \& 220 GHz\\
            Frequency range 150 GHz \dotfill  &[131-169] GHz\\
            Frequency range 220 GHz \dotfill  &[192.5-247.5] GHz\\
            Window Aperture [m]\dotfill  & 0.56 \\
            Focal plane temp. [mK]\dotfill  &300\\
            Sky Coverage\dotfill  &1.5\%\\
            FWHM [degrees]\dotfill &0.39 (150 GHz), 0.27 (220 GHz)\\
        \end{tabular}
    
    \caption{\label{tab:qubic_params}QUBIC main parameters (from Hamilton \textit{et al.} \cite{2020.QUBIC.PAPER1})}   
    \end{center}
\end{table}

Each pair of pupils, called a baseline, contributes with an interference fringe in the focal plane. The whole set of pupils produces a complex interference pattern that we call the synthesized image (or dirty image) of the source. For a multiplying interferometer, the observables are the visibilities associated with each baseline. In bolometric interferometry, the observable is the synthesized image, which can be seen as the image of the inverse Fourier transform of the visibilities.

\begin{figure}[ht!]
      \centering
       \includegraphics[width=0.45\hsize]{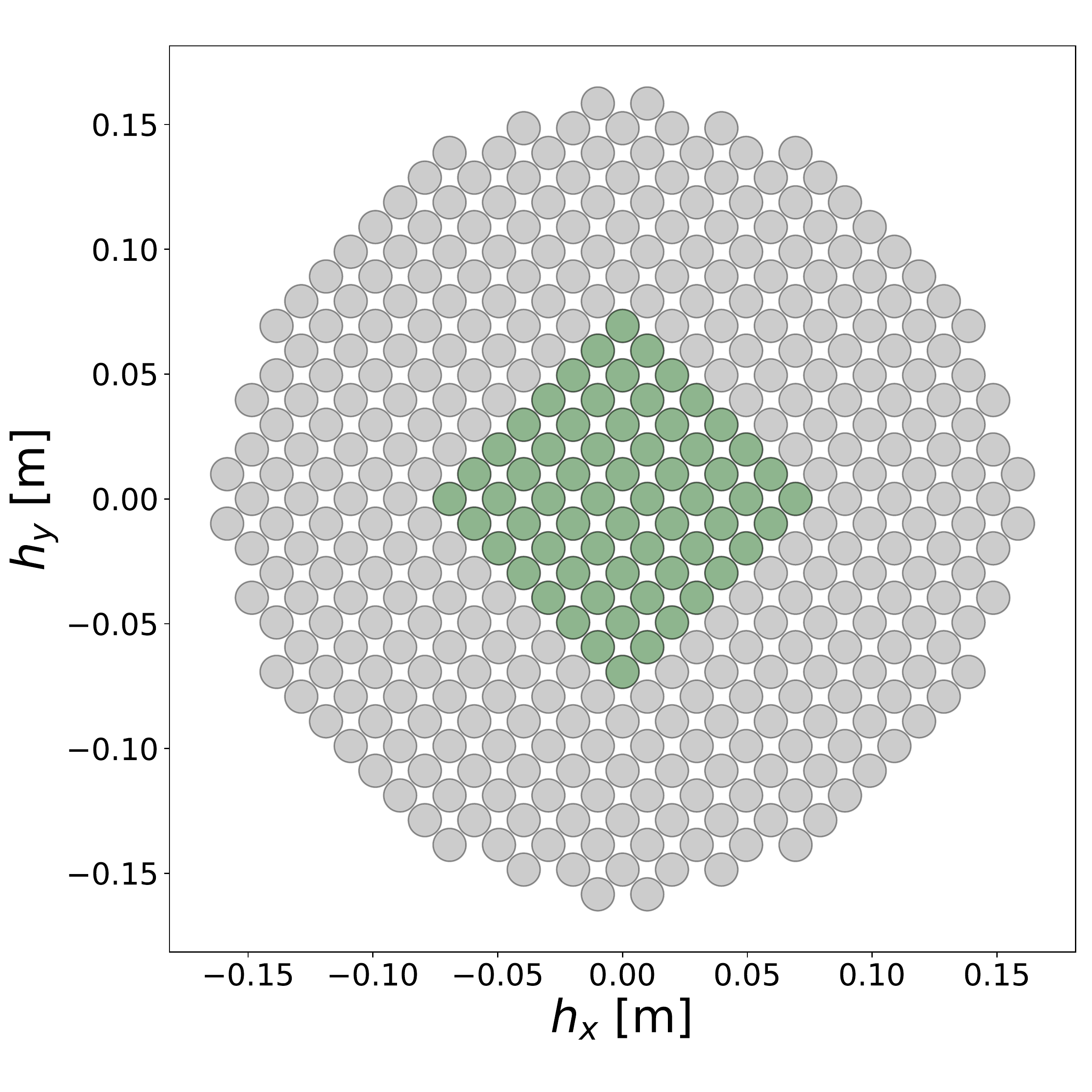}
       \includegraphics[width=0.45\hsize]{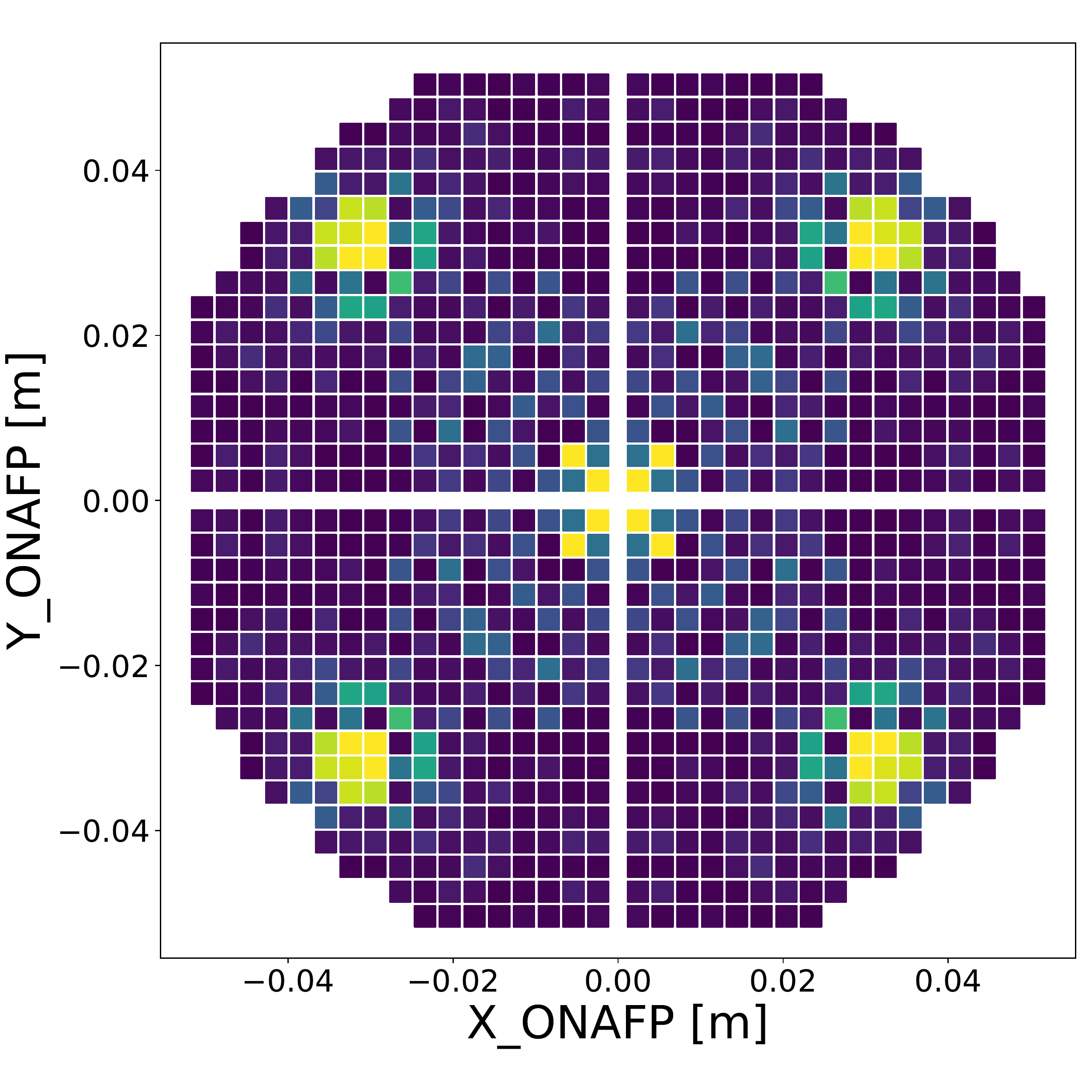}
       \caption{\textit{Left:} Feedhorn array of the FI made of 400~pupils arranged on a rectangular grid $22 \times 22$ within a circle. The TD horn array is highlighted in green. \textit{Right:} Intensity pattern on the focal plane of the FI composed by 992 bolometers for an on-axis point source in the far field emitting at 150~GHz with all horns open. The color scale in intensity is arbitrary.  
       }
        \label{fig:fig_beamFP}
\end{figure}

The left panel of figure~\ref{fig:fig_beamFP} shows the array of back-to-back feedhorns of the full instrument (FI). The feedhorn array is made of 400~pupils arranged on a rectangular grid $22 \times 22$ within a circle. More details on measurements and realistic simulations are presented in Cavaliere \textit{et al.}~\cite{2020.QUBIC.PAPER7} and O'Sullivan \textit{et al.}~\cite{2020.QUBIC.PAPER8}, especially about cross-polarization. However, in this paper, we consider an ideal instrument. The beam looking at the sky is called the primary beam, and secondary beam is the one looking toward the focal plane. In the right panel, we can see the synthesized image obtained on the focal plane of the FI, composed by 992 bolometers, when the instrument is looking at a point source located on the optical axis in the far field.

\subsection{Synthesized imaging}\label{ss:synth_imaging}
A bolometer is a total power detector. The signal $S_\eta(\textbf{r},\lambda)$ on a point $\textbf{r} = (x,y)$ of the focal plane\footnote{$\textbf{r}=0$ at the optical center of the focal plane.} is the square modulus of the electric field $E_\eta(t, \textbf{n},\lambda)$ averaged over time $t$ and integrated over all sky directions $\textbf{n}$. The signal with polarization $\eta$ from each direction is re-emitted by each of the pupils resulting in a path difference in the optical combiner. The  signal on the focal plane depends on the location of each pupil $\textbf{h}_j$, its primary beam $B_\mathrm{prim}(\textbf{n},\lambda)$, the focal length of the combiner $f$, the secondary beam of the pupil on the focal plane $B_\mathrm{sec}(\textbf{r},\lambda)$, and the wavelength $\lambda$:
\begin{eqnarray}
    S_\eta(\textbf{r},\lambda) &=& \int B_\mathrm{prim}(\textbf{n},\lambda) B_\mathrm{sec}(\textbf{r},\lambda) \left<\left| \sum_j E_\eta(t, \textbf{n},\lambda)  \right. \right. \nonumber \\
    &&~~~~~~~\times \left. \left. 
    \exp\left[i 2 \pi \frac{\textbf{h}_j}{\lambda}\cdot\left(\frac{\textbf{r}}{\sqrt{f^2+r^2}}-\textbf{n}\right)\right]\right|^2\right>\mathrm{d}\textbf{n},
    \label{eq_init}
\end{eqnarray}
where $r$ is the norm of $\textbf{r}$. We use the concept of the point spread function (PSF) of the synthesized beam, 
\begin{equation}\label{eq_sb}
    PSF(\textbf{n},\textbf{r},\lambda) =  B_\mathrm{prim}(\textbf{n},\lambda)B_\mathrm{sec}(\textbf{r},\lambda)    \times\left| \sum_j  \exp\left[i 2 \pi \frac{\textbf{h}_j}{\lambda}\cdot\left(\frac{\textbf{r}}{\sqrt{f^2+r^2}}-\textbf{n}\right)\right]\right|^2,
\end{equation}
to rewrite equation~\ref{eq_init} as
\begin{equation}\label{eq_sigPSF}
    S_\eta(\textbf{r},\lambda) = 
    \int \left<\left| E_\eta(t,\textbf{n},\lambda) \right|^2\right> \times 
    PSF(\textbf{n},\textbf{r},\lambda)
    ~\mathrm{d}\textbf{n}.
\end{equation}

The rotating HWP modulates the polarized signal, with a varying angle $\phi$, so we can write equation~\ref{eq_sigPSF} in terms of the synthesized images of the Stokes parameters:
\begin{equation}\label{eq_signal}
    S(\textbf{r},\lambda) = S_I(\textbf{r},\lambda) + \cos(4\phi) S_Q(\textbf{r},\lambda) + \sin(4\phi) S_U(\textbf{r},\lambda)
\end{equation}
where the synthesized images are the convolution of the sky through the synthesized beam:
\begin{equation}\label{conv_sky}
    S_X(\textbf{r},\lambda) = \int X(\textbf{n},\lambda) \times 
    PSF(\textbf{n},\textbf{r},\lambda) ~ \mathrm{d}\textbf{n},
\end{equation}
$X$ standing for the Stokes parameters I, Q or U.
The signal received in the detectors with a bolometric interferometer is therefore exactly similar to that of a standard imager: the sky convolved with a beam. The only difference being that this beam is not that of the primary aperture system (telescope in the case of an ordinary imager) but is given by the geometry of the input pupil array and the beam of the pupils (see equation~\ref{eq_sb}). With such an instrument, one can scan the sky in the usual manner with the synthesized beam, gathering time-ordered-data (TOD) for each sky direction (and orientation of the instrument) and reproject this data onto a map at the data analysis stage (see section~\ref{ss:mapmaking}).

\subsection{Realistic case}\label{ss:real_case}

Note that in a real detector the signal is integrated over the wavelength range defined by filters and also over the surface of the detectors~\cite{Battistelli2011}. If one assumes that the sky signal does not vary within the wavelength range~\footnote{This clarification is only made so that we can write the integral analytically. In the code, the integration is done numerically over small sub-bands ( around 16~sub-bands into the wide band.)}, the expression for the signal, equation~\ref{eq_signal}, is unchanged and one just needs to redefine the synthesized beam as
\begin{eqnarray}
    PSF(\textbf{n},\textbf{r}_d,\lambda_k)=\int\int 
    PSF(\textbf{n},\textbf{r},\lambda) J_{\lambda_k}(\lambda)\Theta(\textbf{r}-\textbf{r}_d) ~\mathrm{d}\lambda \mathrm{d}\textbf{r}
\end{eqnarray}
where $J_{\lambda_k}(\lambda)$ is the shape of the filter for the band centered in $\lambda_k$ and $\Theta(\textbf{r})$ represents the top-hat function for integrating over the detector whose center is at $\textbf{r}_d$.

\subsection{The monochromatic synthesized beam}\label{ss:shape_sb}

\begin{figure}[t]
       \centering
       \includegraphics[width=0.7 \hsize]{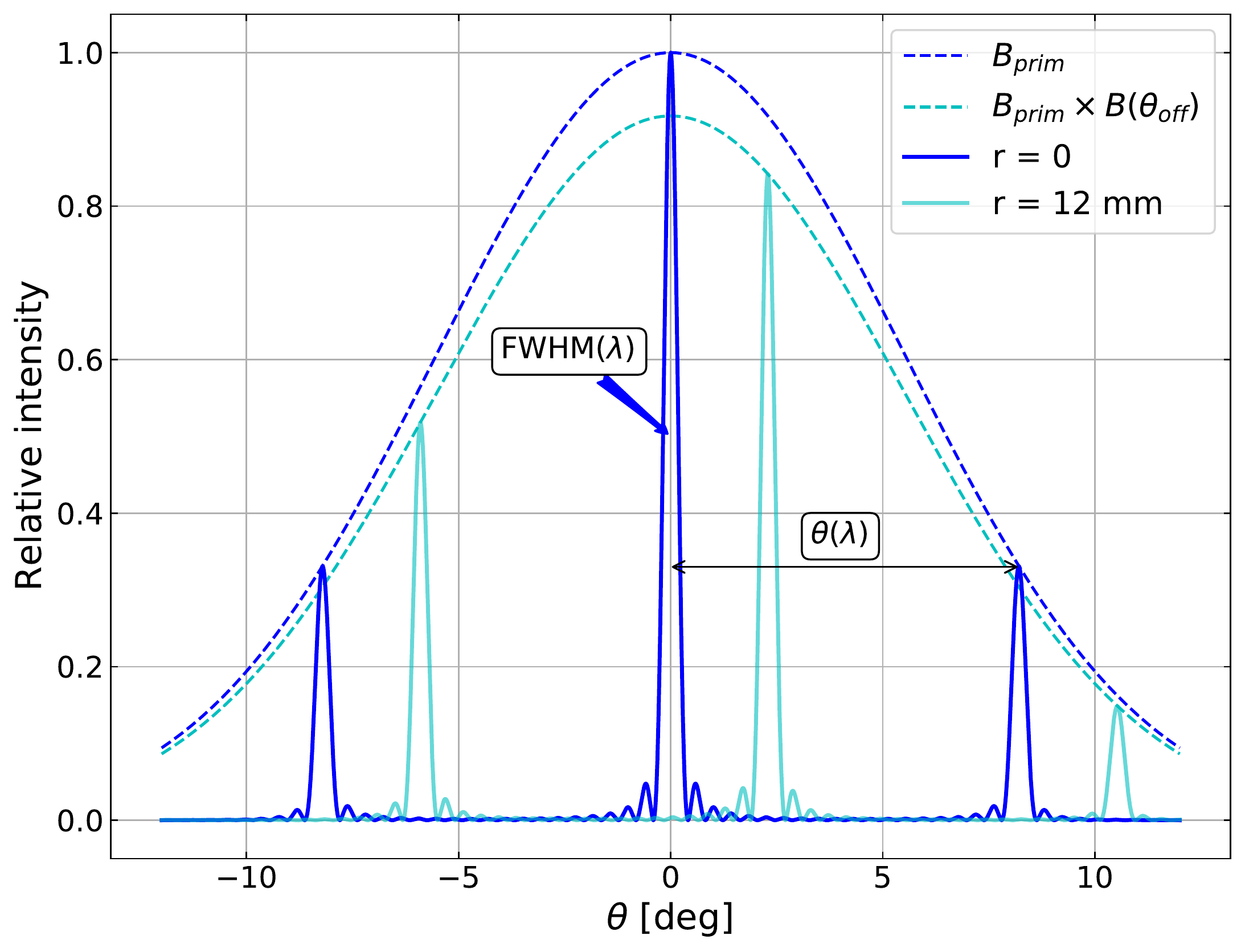}
       \caption{Cut of the synthesized beam as a function of $\theta$ (angle between $\bf{n}$ and the optical axis) given by equation~\ref{approx_sb} for a square array of $20\times 20$ pupils separated by $\Delta h = 14$~mm at 150~GHz frequency (2~mm wavelength) for a detector located at the center of the optical axis in blue and 12~mm apart in cyan. Dashed lines represent the primary beam of the pupils (here Gaussian). Resolution and peak separation depend linearly on the wavelength $\lambda$.
       }
        \label{fig:fig_sb}
    \end{figure}

If the pupil array is a regular square grid of $P$ pupils on a side spaced by a distance $\Delta h$, the sum in equation~\ref{eq_sb} can be analytically calculated (see O'Sullivan \textit{et al.}~\cite{2020.QUBIC.PAPER8} for more detail) and the monochromatic point-like synthesized beam, assuming $f$ is large enough to use the small-angle approximation, becomes
\begin{equation}
    PSF(\textbf{n},\textbf{r},\lambda) = B_\mathrm{prim}(\textbf{n}, \lambda) B_\mathrm{sec}(\textbf{r}, \lambda) 
    \times
    \frac{
        \sin^2\left[P \pi \frac{\Delta h}{\lambda}\left( \frac{x}{f}-n_x\right)     \right]
        }
        {
        \sin^2\left[\pi \frac{\Delta h}{\lambda}\left( \frac{x}{f}-n_x\right)     \right]
        }
    \frac{
        \sin^2\left[P \pi \frac{\Delta h}{\lambda}\left( \frac{y}{f}-n_y\right)     \right]
        }
        {
        \sin^2\left[\pi \frac{\Delta h}{\lambda}\left( \frac{y}{f}-n_y\right)     \right]
        }.
    \label{approx_sb}
\end{equation}
where $\textbf{n} = (n_x, n_y)$ is the off-axis angle of the source. In such a case, the synthesized beam of the monochromatic point-like detectors has the shape of a series of large peaks with ripples in between, modulated by the primary beam of the pupils. For illustrative purpose, a cut of the synthesized beam is shown in figure~\ref{fig:fig_sb} for such a square $20\times20$ array of horns. The realistic synthesized beam corresponding to our circular array (see figure~\ref{fig:fig_beamFP}) is shown by O'Sullivan \textit{et al.}~\cite{2020.QUBIC.PAPER8} (figure 11) and shows minor differences.
Figure~\ref{fig:fig_sb} shows this approximate synthesized beam for two different detectors emphasizing the fact that the location of the peaks moves with the detector location in the focal plane. Moreover, the intensity received by the detector changes inversely with $r$ so the two Gaussian envelopes also differ. From the expression in equation~\ref{approx_sb} it is straightforward to see that the full width at half maximum (FWHM) of the large peaks is roughly given by $\mathrm{FWHM}=\frac{\lambda}{(P-1)\Delta h}$ while their separation is $\theta=\frac{\lambda}{\Delta h}$ as illustrated in figure~\ref{fig:fig_sb}.

The real synthesized beam of the QUBIC TD has been measured in the lab using a monochromatic calibration source and is presented by Torchinsky \textit{et al.}~\cite{2020.QUBIC.PAPER3}. For the TD, the horn array is an $8\times8$ square array. The measured synthesized beam is in overall good agreement with the approximate analytical expression in equation~\ref{approx_sb} and figure~\ref{fig:fig_sb} although excursions from this perfect case are expected due to optical defects and diffraction in the optical chain. It was shown by Bigot-Sazy \textit{et al.}~\cite{bigot2013self} that the shape of the synthesized beam for each detector can be precisely recovered through the ``self-calibration'' technique that is heavily inspired from synthesis imaging techniques~\cite{cornwell1981new}.

\subsection{Monochromatic map-making}\label{ss:mapmaking}
Before discussing spectral imaging, we first describe the map-making with a bolometric interferometer in the monochromatic case. As shown in section~\ref{ss:synth_imaging}, the instrument is essentially equivalent to a standard imager, scanning the sky with the synthesized beam, producing TOD that can be projected onto sky maps. The map-making will therefore be very similar to that of a standard imager. In the monochromatic case, if the sky signal is $\vec{s}$, the TOD $\vec{y}$ can be written as:
\begin{equation}
    \vec{y}=H\cdot \vec{s} +\vec{n}
\label{eq:TOD_1band}
\end{equation}
where $\vec{n}$ is the noise and $H$ is an operator that describes the convolution by the synthesized beam in equation~\ref{conv_sky} as well as the pointing at the different directions of the sky. $H$ is a 2-dimensional matrix: number of time samples (scaling with the number of detectors) $\times$ number of sky pixels. The noise has two contributions: photon noise and detector noise. Photon noise is the Poisson fluctuations from the temperature of the CMB (T$_{\rm CMB} \simeq 2.7$~K), the atmosphere, and the internal optical components. Detector noise is given by the noise equivalent power (NEP) measured in each detector. The atmosphere emissivity is assumed to be stable, we did not consider the effect of fluctuations in the atmospheric load, this requires a significant step forward  in  our  pipeline that  we  are  currently  carrying  out. Equation~\ref{eq:TOD_1band} will be generalized to the case of several frequency sub-bands in section~\ref{ss:SI-capabilities}. 

\begin{figure}[t]
   \centering\resizebox{\hsize}{!}{
   \includegraphics{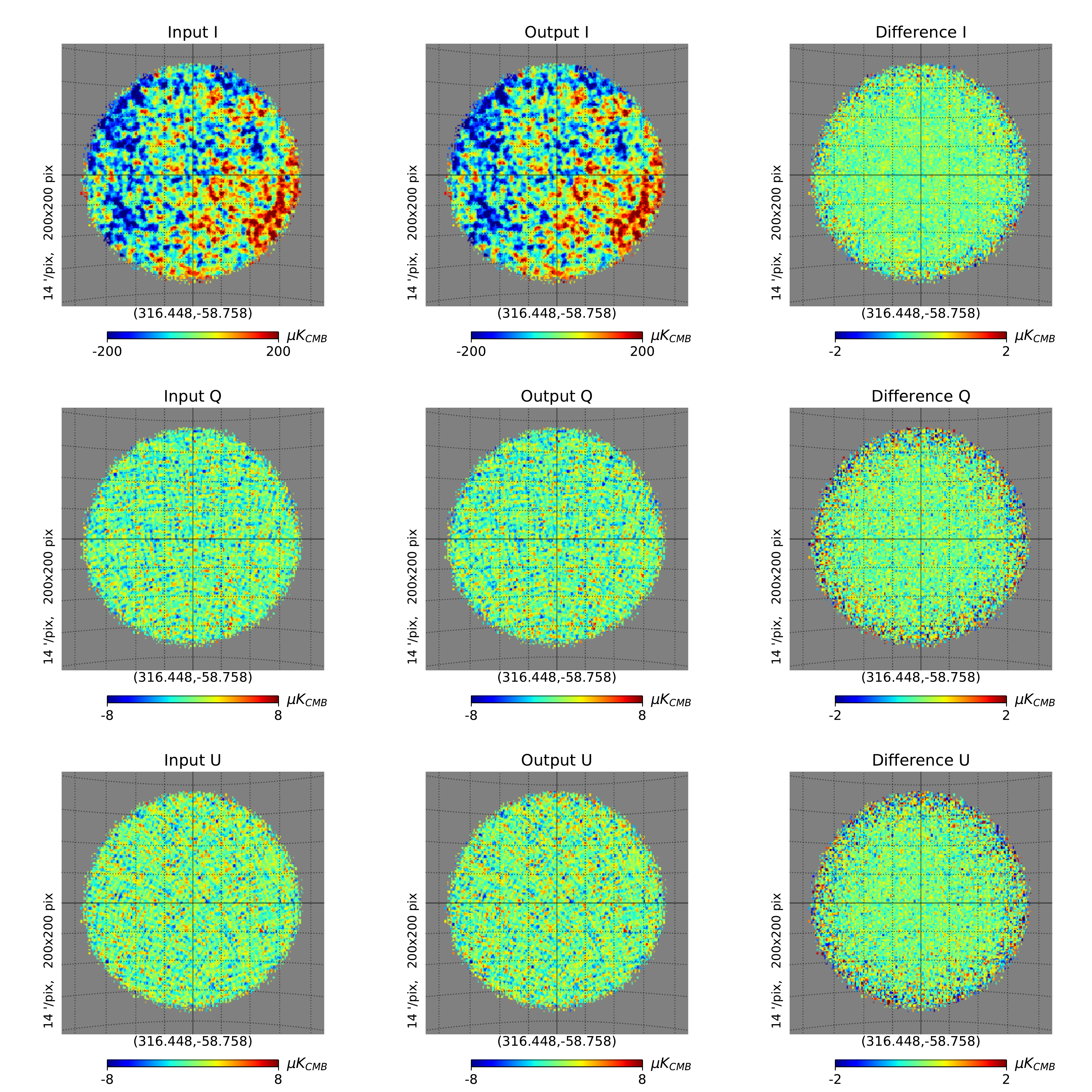}}
   \caption{Result of the map-making for IQU Stokes parameters for a bolometric interferometer pointing in a 15~degree radius sky patch containing only CMB. The first column is the input sky convolved at the resolution of the instrument using a Gaussian with a FWHM equal to 0.4~degrees. The second column is the sky reconstructed by map-making. The last column is the difference between both. The grid lines are equally spaced every 5$^\circ$. This simulation was obtained with the QUBIC full pipeline in the 220~GHz band. The noise was scaled to 4 years of observations. This simulation required 127 iterations to converge.}
    \label{fig:fig_mm}
\end{figure}

For standard imagers, the $H$ operator is such that each line (corresponding to sky pixels contributing to one time sample) only contains a single non-zero value, meaning that $\vec{s}$ is actually the sky map convolved to the instrument's resolution and that the instrument samples the convolved sky with a single peak \cite{tegmark1997cmb, borrill1999madcap}. 
    
In the case of a bolometric interferometer, this assumption is not valid due to the multiple peaked shape of the synthesized beam (see figure~\ref{fig:fig_sb}) which makes it impossible to use the map-making algorithms usually developed for direct imagers. We use instead an algorithm that starts from an initial guess and then simulates iterative maps $\vec{s}_i$, where $i$ is the iterative index\footnote{The software uses the massively parallel libraries~\cite{chanial2012pyoperators} developed by P. Chanial pyoperators (\href{https://pchanial.github.io/pyoperators/}{https://pchanial.github.io/pyoperators/}) and pysimulators (\href{https://pchanial.github.io/pysimulators/}{https://pchanial.github.io/pysimulators/}).}. For each of these maps, we apply the bolometric interferometer acquisition model, taking into account the scanning strategy of the sky, and we construct TOD $\vec{y}_i$ that are then compared to the data TOD $\vec{y}$ using a merit function that accounts for the noise in the TOD domain. In the case of stationary and Gaussian distributed noise, the maximum likelihood solution is reached by minimizing the $\chi^2$:
\begin{equation}
    \chi^2(\vec{s}_i) = \left(\vec{y}-\vec{y}_i\right)^T\cdot N^{-1}\cdot \left(\vec{y}-\vec{y}_i\right)
\label{eq_chi2}
\end{equation}
where $N$ is the covariance matrix of the noise. We minimize equation~\ref{eq_chi2} to find the best simulated sky map, $\hat{\vec{s}}$, using a preconditioned conjugate gradient method~\cite{hestenes1952methods, shewchuk1994introduction}. This is jointly done for the IQU Stokes parameters and results in unbiased estimates of the maps as shown in figure~\ref{fig:fig_mm}.

\section{Spectral dependence}\label{sec:specSection}

\subsection{The polychromatic synthesized beam}\label{ss:poly-sb}
As can be seen in equation~\ref{approx_sb}, the synthesized beam is directly dependent on wavelength and this is shown in figure~\ref{fig:fig_sb}.  

\begin{figure}[t]
    \centering
    \includegraphics[width=1 \hsize]{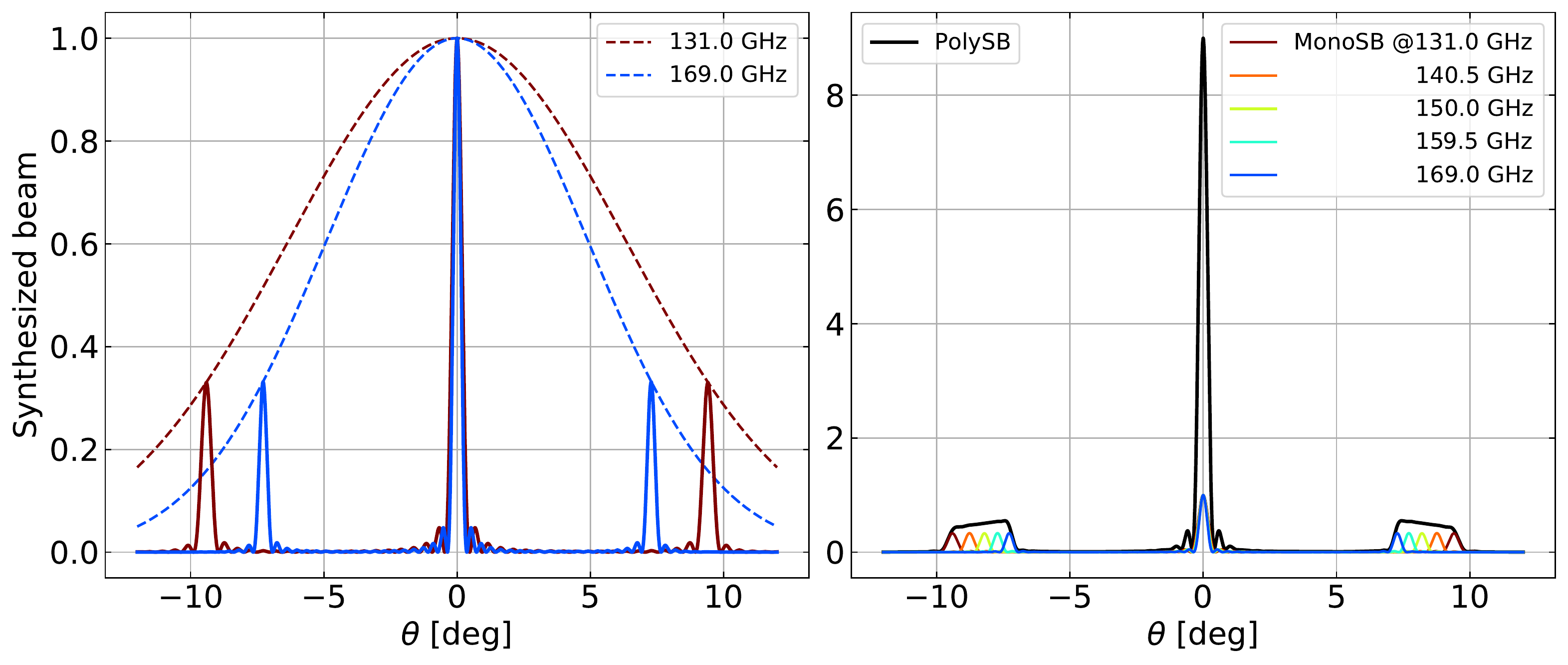}
    \caption{\textit{Left:} Monochromatic synthesized beam (\textit{MonoSB}) for 131~and 169~GHz. Each synthesized beam is modeled according to equation~\ref{approx_sb} for a square array of $20 \times 20$ pupils separated 14~mm apart at 131~and 169~GHz. The primary beam at each frequency is shown by a dashed line.  \textit{Right:}~Polychromatic beam (\textit{PolySB}, black line) as result of the addition of 9~monochromatic synthesized beams (5~of them are shown in colored lines) spanning our 150~GHz band (131 to 169~GHz). We sample the continuous frequency band with discrete frequencies.}
    \label{fig:fig_sb_freq}
\end{figure}

The off-axis angle (given by the primary beam of the pupils), the FWHM of the peaks (hence the resolution of the maps), and the angle on the sky between two peaks all depend linearly on $\lambda$. This dependence on wavelength can be exploited to achieve spectral imaging capabilities. Within a wide band, the synthesized beam will be the integral of the synthesized beam of all the monochromatic contributions within the band resulting in a polychromatic synthesized beam. Figure~\ref{fig:fig_sb_freq} shows the cross cut of the synthesized beams modeled according to equation~\ref{approx_sb}. The left panel shows the monochromatic synthesized beam for 131~and 169~GHz while the right panel shows a polychromatic synthesized beam using 9~monochromatic synthesized beams. 

With a bolometric interferometer operating over a large bandwidth, for each pointing towards a given direction in the sky, one gets contributions from all the multiple peaks in the synthesized beam at all frequencies.  As a result, we have both spatial and spectral information in the TOD.  Precise knowledge of the synthesized beam along the frequency will then allow one to reconstruct the position and amplitude of the sky in multiple frequency sub-bands.

\subsection{Spectral imaging capabilities}\label{ss:SI-capabilities}
The synthesized beam at two different frequencies $\nu_1$ and $\nu_2$ will be distinguishable from one another as long as their peaks are sufficiently separated.
The angular separation between the two peaks $\Delta \theta= \frac{c \Delta \nu}{\nu^2 \Delta h}$ (where $\Delta\nu=\nu_2-\nu_1$ and $\nu=\sqrt{\nu_1\nu_2}$) must be large enough to unambiguously distinguish the two peaks. In order to have an order of magnitude of the possible width for each sub-band we consider a square array of pupils and we apply the Rayleigh criterion~\cite{rayleigh1879xxxi}:  
\begin{eqnarray}
    \frac{c \Delta \nu}{\nu^2 \Delta h} \gtrsim \frac{c}{\nu(P-1)\Delta h} 
    \quad \Leftrightarrow \quad 
    \Delta\nu \gtrsim \frac{\nu}{(P-1)}
\label{eq:Rayleigh_criteria}
\end{eqnarray}
where $P$ is the number of pupils on a side of the square-packed pupil-array. A bolometric interferometer therefore not only has a resolution on the sky $\mathrm{FWHM}_\theta \simeq \frac{c}{\nu (P-1) \Delta h}$, but also in electromagnetic frequency space $\frac{\Delta\nu}{\nu}\simeq\frac{1}{P-1}$. 

The map-making presented in section~\ref{ss:mapmaking} can be extended in order to build, simultaneously, with the same TOD, maps at a number of different frequencies as long as they comply with the frequency separation given above. The iterative TOD $\vec{y_i}$ can be written as:
\begin{equation}\label{eqmmfreq}
    \vec{y_i} = \sum_{j=0}^{N_{\rm rec}-1} H_j \hat{\vec{s}}_{ij} + \vec{n}
\end{equation}
where $H_j$ describes the acquisition (convolution$+$pointing) operator with the synthesized beam at frequency $\nu_j$, $\hat{\vec{s}}_{ij}$ is the sky signal estimator at iteration $i$ for the frequency $\nu_j$ and $N_{\rm rec}$ is the number of reconstructed sub-bands. Similarly, as in the map at a single frequency (figure~\ref{fig:fig_mm}), one can recover the maps $\vec{s_j}$ by solving equation~(\ref{eqmmfreq}) using a preconditioned conjugate-gradient method (see section~\ref{sec:testSection} for corresponding simulations). Thus, the instrument acquires data in a single wide frequency band and spectral imaging technique applies during map-making, in the post-processing of the data.

The QUBIC FI has two wide-bands centered at 150 and 220~GHz with $\Delta\nu/\nu=0.25$
and a 400-feedhorn array
packed on a square grid within a circular area (see figure~\ref{fig:fig_beamFP}). We approximate it with a square grid $20\times 20$ so $\frac{1}{(P-1)}\sim 0.05$. It is thus possible to reconstruct approximately 5~sub-bands in each of the initial bands of QUBIC. Note that this number should just be taken as an order of magnitude for the achievable number of sub-bands but we will see in figure~\ref{fig:resolutions} that it is a good approximation. In section~\ref{sec:noiseSection} we will show that reconstructing up to 8 sub-bands is feasible but with a significant degradation of the signal-to-noise ratio.

\section{Testing spectral imaging on simple cases}\label{sec:testSection}
We can use the QUBIC simulation pipeline to test the spectral imaging capabilities of bolometric interferometry in a simple way. Some concepts and parameters used in simulations are defined in table~\ref{tab:parameters}.

\begin{table}[t]
    \begin{center}
    \begin{minipage}{\textwidth}
        \begin{tabular}{l p{12cm} p{8cm}} 
            \hline
            Parameter & Details \\
            \hline
            \hline
            $N_{\rm in}$ & Number of input or true maps (in $\mu$K) used to simulate a broadband observation (TOD). Each map represents a sky at a specific frequency $\nu_j$ for the IQU Stokes parameters\footnote{Skies are generated using PySM: Python Sky Model~\cite{thorne2017python}.} \footnote{Maps are projected using HEALPix: Hierarchical Equal Area Isolatitude Pixellization of sphere~\cite{gorski2005healpix}.}. Possible values: 15, 16, 18, 21 or 48.   \\
            
            $N_{\rm rec}$ & Number of sub-bands reconstructed from a single broadband observation, from 1 to 8. In all simulations $N_{\rm rec}$ is a divisor of $N_{\rm in}$.\\
            
            $N_{\rm conv}$ & Number of convolved maps equal to $N_{\rm rec}$. Each of these maps is obtained convolving the $N_{\rm in}$ input maps at the QUBIC spatial resolution corresponding to that input frequency  and then averaging within the reconstructed sub-band. \\
            \hline
            $\theta$ & Radius of sky patch observed in simulations. Value: 15~degrees. \\
            
            Pointings & Number of times that the instrument observes in a given sky direction aligned with the optical axis. Values $ > 10^{4}$.\\
            \hline
            NEP$_{\rm det}$ & Detector noise equivalent power (NEP) added as white noise. Value: $4.7\times10^{-17}$~W/$\sqrt{\rm Hz}$.\\
            
            NEP$_{\gamma}$ & Photon NEP added as white noise in time-domain, calculated from the atmospheric emissivity measured in our site, as well as emissivities from all components in the optical chain. The atmospheric load is assumed to be stable. The value is different for each detector because of their different illumination by the secondary beam $B_{sec}$.
            The average value at 150~GHz is $4.55\times 10^{-17}$~W/$\sqrt{\rm Hz}$ and $1.72\times 10^{-16}$~W/$\sqrt{\rm Hz}$ at 220~GHz.\\
            \hline
            \hline
        \end{tabular}
        \end{minipage}
        \caption{\label{tab:parameters} Typical parameters used in acquisition, instrument and map-making to do an end-to-end simulation. A preconditionned conjugate gradient method is used for map-making. }
    \end{center}
\end{table}

\subsection{Extended source reconstruction}\label{ss:ext_source}

\begin{figure}[t]
    \centering\resizebox{1.\hsize}{!}{\centering{\includegraphics{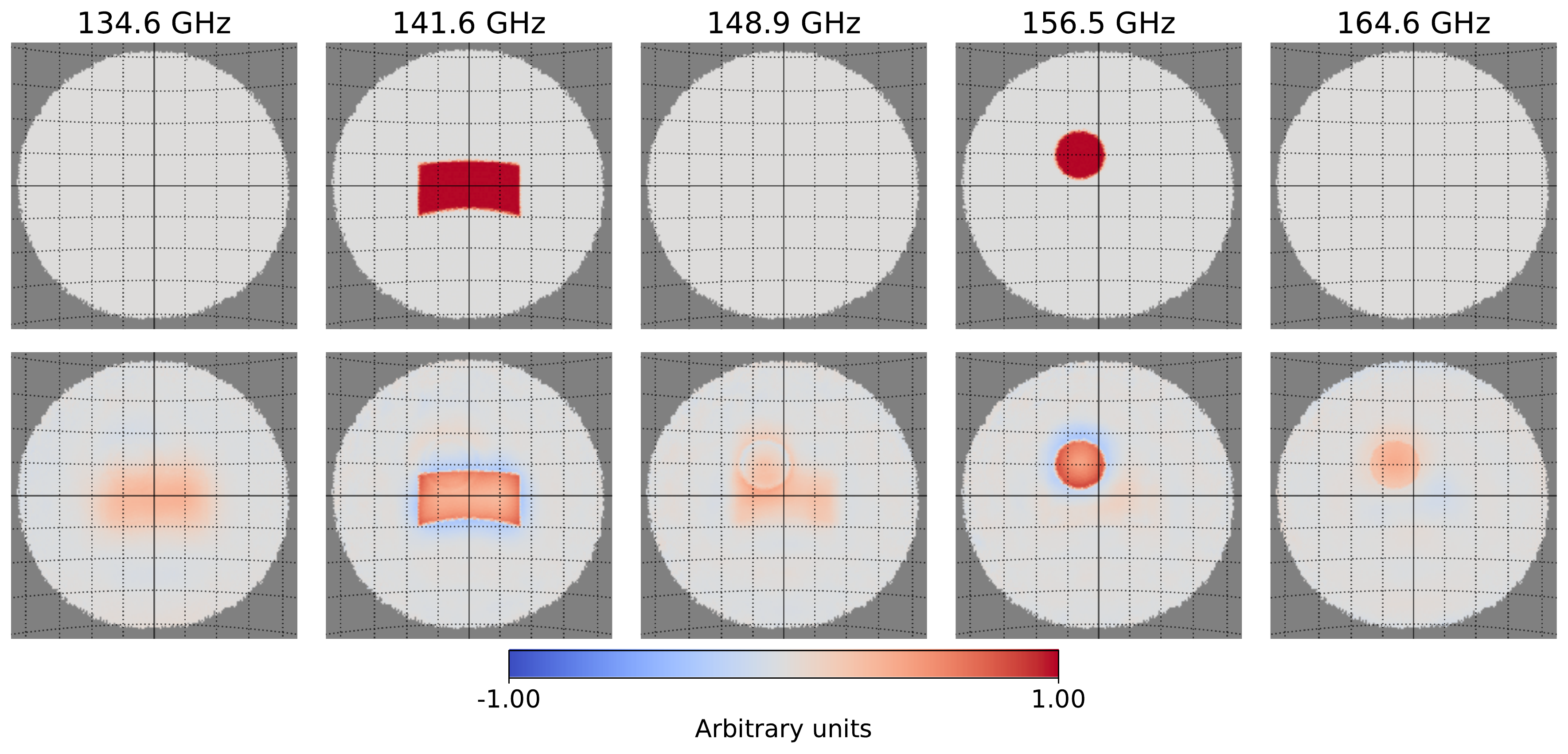}}}    \caption{Map-making for a sky full of zeros with two extended, monochromatic regions centered at $141.6$ (square) and $156.5$~GHz (disk). Each column corresponds to one sub-band. The first row shows the input sky maps spatially convolved at the QUBIC resolution. The reconstructed maps using the QUBIC pipeline are shown in the second row. The units are arbitrary. 
    }
\label{fig:ext-sources}
\end{figure}

The input map used for this example to simulate TOD is composed of zeros in each pixel of its $15$ input frequencies, $N_{\rm in}$, and for the three Stokes components. Two monochromatic extended regions with a high signal-to-noise ratio are added: a square centered at $141.6$~GHz and a disk centered at $156.5$~GHz. Map-making is done for 5~sub-bands centered at $134.6$, $141.6$, $148.9$, $156.5$ and $164.6$~GHz, with a bandwidth of $6.8, 7.1, 7.5, 7.9 $ and $8.3$~GHz respectively. The scan is performed with $8500$ points randomly placed over a 150~square degrees sky patch. Noise is included in the TOD.

The first row in figure~\ref{fig:ext-sources} shows the input sky maps spatially convolved at the QUBIC resolution. 
The second row shows the reconstructed maps after map-making onto five sub-bands. The maps are normalized by the maximum value in the convolved maps. In the first, third and fifth sub-bands, where originally the signal is zero, structures corresponding to the signals of neighboring sub-bands appear. The median of the signal on such sub-bands are $0.27$, $0.19$, $0.31$ respectively. The medians are calculated within a region defined by the shape of the signal in the adjacent sub-band, i.e. rectangle (1st sub-band), rectangle + disk (3rd) and disk (5th). The leakage is due to the fact that during the map-making process, it occurs from the frequencies where the monochromatic signal was located towards the neighboring sub-bands due to the frequency point spread function (FPSF) that will be studied in section~\ref{ss:fpsf}.

\subsection{Angular resolution}
\label{ss:angular_reso}
\begin{figure}[t]
    \centering\resizebox{0.9\hsize}{!}{\centering{\includegraphics{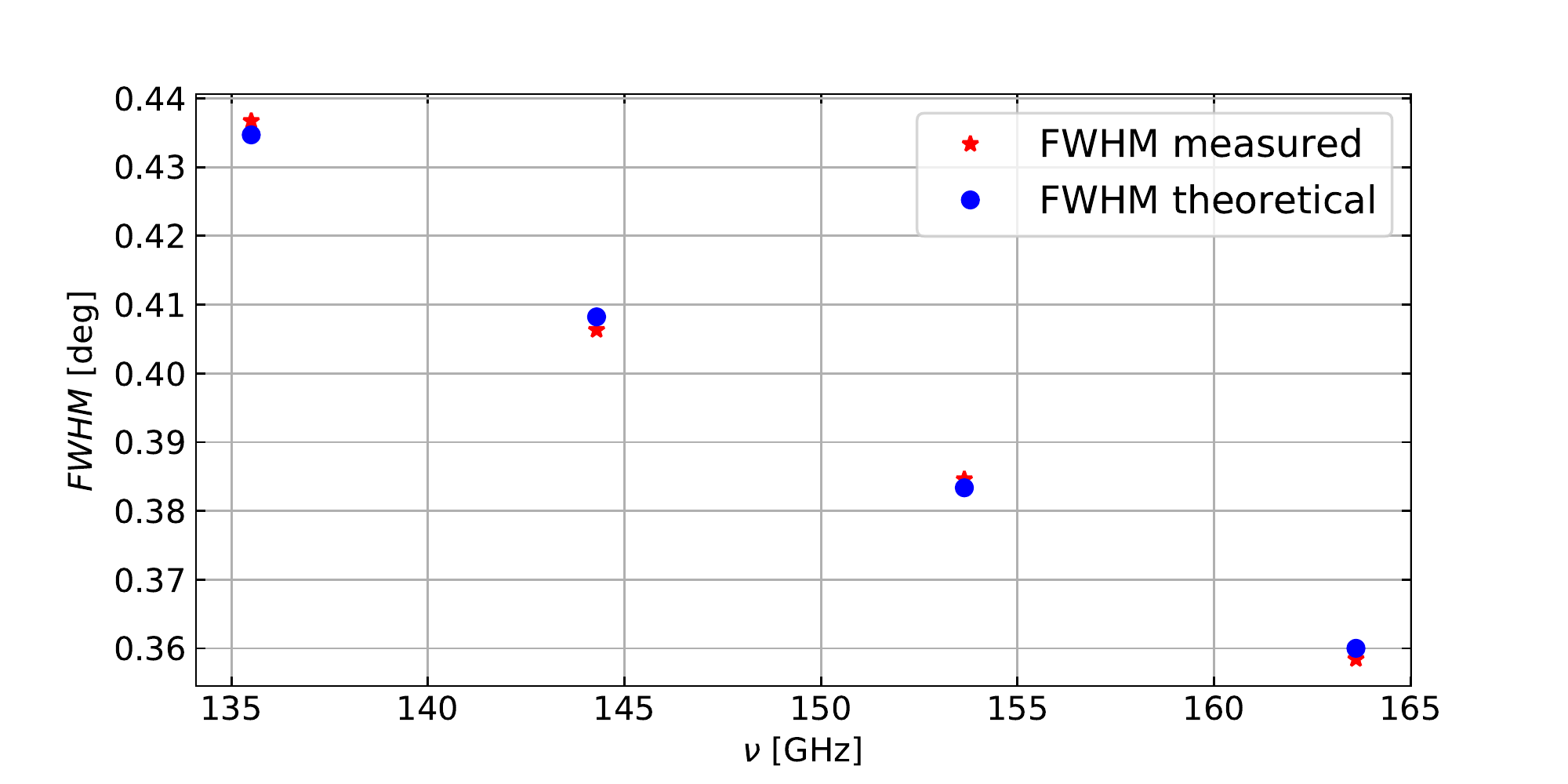}}}
    \caption{Red stars represent the angular resolution measured for a single point source emitting in the broad band after map-making onto 4 sub-bands. Blue dots are the theoretical values expected for each central frequency of the sub-bands.
    }
\label{fig:angular-resolution}
\end{figure}

As an example, we used the end-to-end pipeline to simulate the reconstruction onto 4~sub-bands of a point source emitting with a flat spectrum in the 150~GHz wide band. Figure~\ref{fig:angular-resolution} shows the measured (red stars) and theoretical (blue dots) values of the FWHM at the central frequency of each sub-band. Theoretical values are obtained from a quasi-optical simulation~\cite{2020.QUBIC.PAPER8} at 150~GHz and scaled proportionally to frequency. Measurements were made on HEALPix maps and corrected by pixel size and resolution~\cite{2020BAAA...61B.155G}. The difference between measured and theoretical values are up to $0.5\%$ which makes it acceptable. The real angular resolution will be determined once QUBIC is installed on the site using far-field observations (astronomical objects and/or calibration tower). This analysis was done for a flat spectrum over the wide frequency band but this hypothesis is not required. In the code, it is only assumed to be flat in small frequency sub-bands with bandwidths around 1/20 of the total bandwidth. The result would be similar in case of a non flat spectrum.

\subsection{Frequency point spread function characterization}\label{ss:fpsf}

In section~\ref{ss:ext_source} it was shown that the reconstructed map for a sub-band has a fraction of signal coming from neighboring bands (see figure~\ref{fig:ext-sources}). In order to study this effect, we simulate monochromatic point source reconstruction to characterize the FPSF.

If we consider a monochromatic input signal with a spectral energy density $S_{\rm in}(\nu)$ [W/Hz], and considering ideal map-making, then the intensity of the output map $I_{\rm out}(\nu)$ [W] will be given by the convolution of the input signal with the FPSF [unitless]:
\begin{equation}\label{eq:frequency-res}
I_{\rm out}(\nu) = \left[S_{\rm in} \otimes \mathrm{FPSF}\right] (\nu).
\end{equation}
Thus, for a monochromatic source at $\nu_{\rm in}$ with spectral energy density $S_{\rm in}(\nu) = I_0\delta (\nu - \nu_{\rm in})$, we can obtain the FPSF by measuring the intensity in the reconstructed sub-bands.

\begin{figure}[t]
    \centering\resizebox{0.85\hsize}{!}{\centering{
    \includegraphics{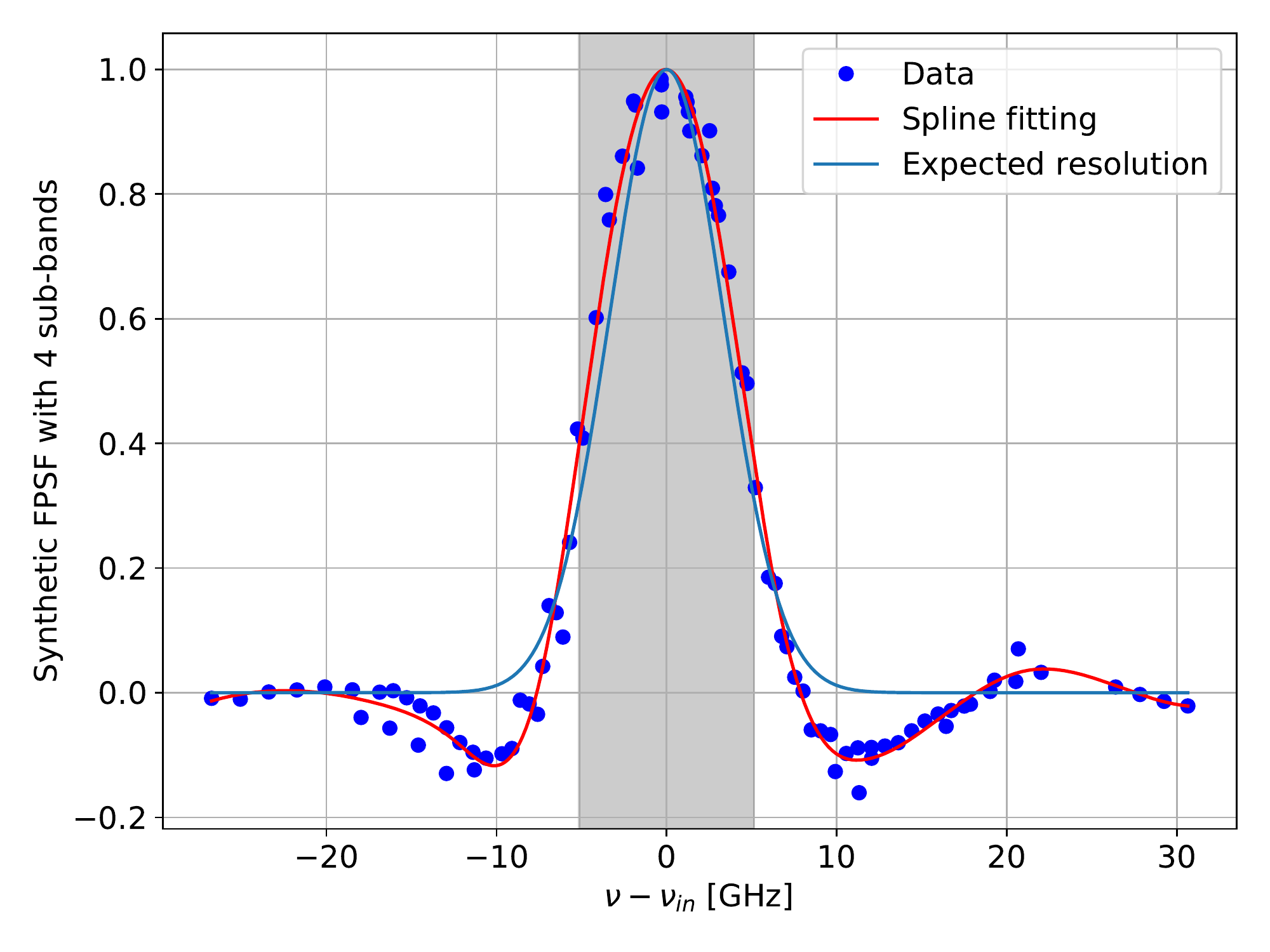}}}    
    \caption{Synthetic FPSF when reconstructing in 4 sub-bands. Blue dots are the intensity of the central pixel of the point source in each sub-band and each simulation superimposed (see text for detailed explanation). Red line is a spline fit. Blue line is a Gaussian with $\textrm{FWHM} = \frac{\nu_0}{P-1}$, $P=20$ and $\nu_0 = 150$~GHz, showing an order of magnitude of the expected frequency resolution (see section~\ref{ss:SI-capabilities}). Strong-grey band represent a typical width of a sub-band when reconstruction is performed for 4 sub-bands.}
    \label{fig:resolutions}
\end{figure}

In order to investigate the FPSF of the instrument, we simulate a scan in frequency by moving $\nu_{\rm in}$ in a high-resolution frequency grid. We use a grid with $N_{\rm in} = 48$ which gives us a resolution of $\sim 0.78$~GHz for the 150~GHz wide band (it would be $\sim1.15$~GHz for the 220~GHz band). This grid allows to improve the map within the spectral range and thus obtain more precise information on how the signal is reconstructed at the center and edge of each sub-band. We performed 22~independent simulations of monochromatic point sources with high signal-to-noise ratio. We kept the spatial location of the point source unchanged and we varied its frequency $\nu_{\rm in}$, covering a spectral range from $133$ to $162.25$~GHz. 

We present results for map-making onto 4~sub-bands with central frequencies at $\nu_{c} = 135.5, 144.3, 153.6, 163.6$~GHz and bandwidths of $\Delta\nu_c = 8.5,  9.1,  9.7, 10.3$~GHz respectively. The FPSF depends on the number of sub-bands so this result is specific to 4~sub-bands. For each simulation, we measure the intensity, normalized to the input one, of the central pixel of the point source in each reconstructed sub-band. In the hypothesis that the FPSF does not depend on the input frequency, we can superimpose all the results. This is done in figure~\ref{fig:resolutions} (blue dots) by subtracting $\nu_{\rm in}$ to center all the measurements on $0$. All the points are well superimposed on a continuous line that we will call the synthetic FPSF of the instrument with 4~sub-bands. A spline fitting is shown in figure~\ref{fig:resolutions} (red line) to get a model for the synthetic FPSF. We also plot a Gaussian (blue line) with $\textrm{FWHM} = \frac{\nu_0}{P-1}$, $P=20$ and $\nu_0 = 150$~GHz, showing an order of magnitude of the expected frequency resolution as explained in section~\ref{ss:SI-capabilities}. As expected from figure~\ref{fig:ext-sources}, we observe that the FPSF extends beyond a single sub-frequency and should be accounted for in the data analysis. This means that our reconstructed sub-bands are not independent from each other and we should expect noise correlation between sub-bands. Because the FPSF is negative in the nearest band we should expect the noise correlations to be negative between neighbouring sub-bands. This will be studied in section~\ref{ss:noise_maplevel}. 
    
\subsection{Galactic dust}
\label{section:galaxy_dust}
\begin{figure}[t]
    \centering
    \includegraphics[width=0.8\linewidth]{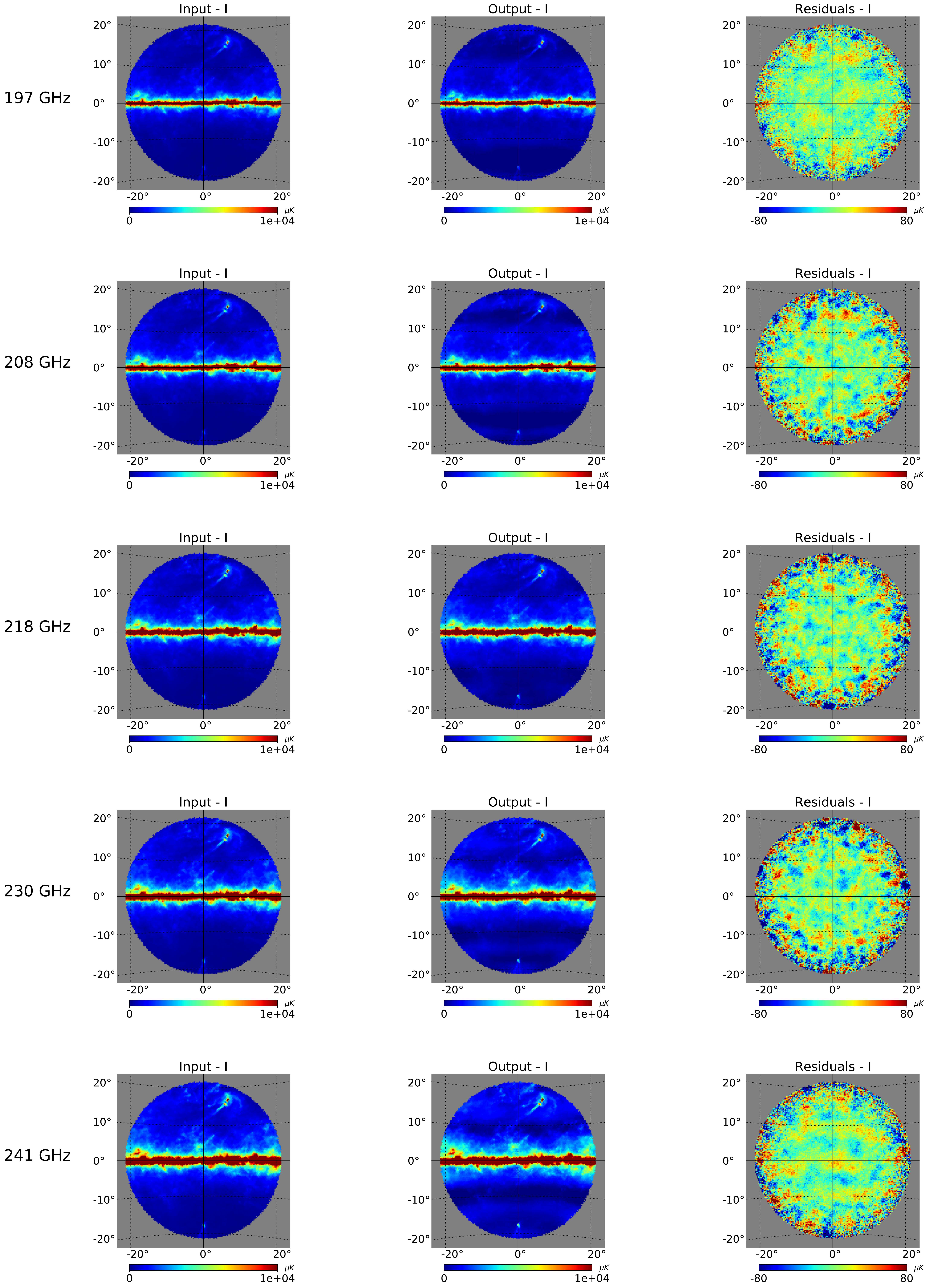}
    \caption{Map-making of the galaxy dust in $N_{\rm rec}=5$~frequency sub-bands from 192 to 247~GHz for I component. The map unit is $\mu$K CMB. The first column is the input sky convolved to the resolution of the instrument in that sub-band. The second column is the reconstructed map after the map-making process. Residuals, defined by the difference between the simulation including noise and a noiseless one, are shown in the last column. The noise is not white but has spatial correlations due to the deconvolution with the multi-peak synthesized beam, it is clearly visible in the 241~GHz sub-band.}
    \label{fig:simu_galaxy_mapI}
\end{figure}

\begin{figure}[t]
    \centering
    \includegraphics[width=0.85\linewidth]{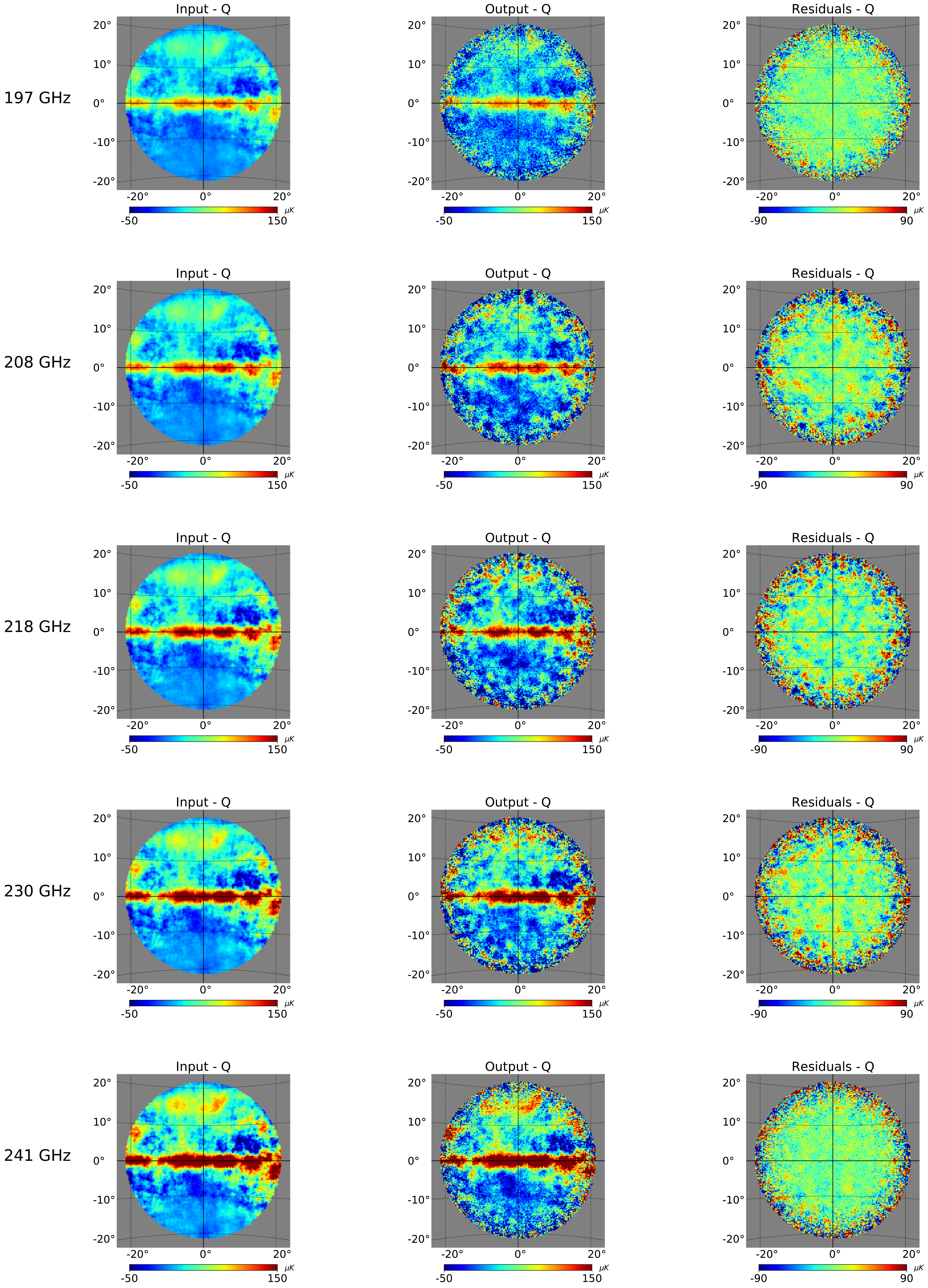}
    \caption{Map-making of the galaxy dust in 5 frequency sub-bands from 192 to 247~GHz. Same as figure~\ref{fig:simu_galaxy_mapI} but for Q component.}
    \label{fig:simu_galaxy_mapQ}
\end{figure}

We demonstrate spectral imaging capabilities by trying to recover the frequency dependence of the dust emission with simulated observations towards the Galactic center. The sky maps contain IQU Stokes parameter components and the dust model is the one provided by PySM3, named \texttt{d1}~\cite{thorne2017python}. We simulate an observation in a sky patch of 15~degree radius. This choice was made in order to minimize the number of pointings required to get a sufficient coverage and so to reduce computational resources needed for simulations. However, a bigger patch would give similar results, even better as border effects would be reduced. The parameters of the pipeline are set in such a way that the simulated instrument has a single focal plane operating either at 150~GHz or at 220~GHz with a 25\% bandwidth each. The wide band TOD are formed through the sum of a number of monochromatic TOD throughout the wide bandwidth as shown in equation~\ref{eqmmfreq}. For this simulation we have used $N_{\rm in}$ (see table~\ref{tab:parameters}) input maps covering the ranges from 137 to 162~GHz and from 192 to 247~GHz. From these wide-band TOD, we are able to reconstruct several numbers of sub-bands using spectral imaging. We have performed simulations with $N_{\rm rec}=1, 2,..., 8$ reconstructed sub-bands. NEP, including photon noise and detector noise (see table~\ref{tab:parameters}) are added as white noise for each TOD. In each case, we perform a Monte-Carlo analysis to get several independent noise realizations and also a noiseless reconstruction that will be the reference. Those end-to-end simulations require high memory usage and need to be parallelized on several machines.\footnote{For instance, with 10000 pointings, 992 detectors (FI), considering the main 14 peaks in the synthesized beam, the size of the pointing matrix for each input sub-band is: $\sim 10000 \times 992 \times 14 \times 16 / 1024^3 \sim 2$~GiB and typically, there are $N_{\rm in} \sim 20$ sub-bands. Regarding the convergence of the map-making, the number of iterations needed will vary, especially with the level of noise and with the number of sub-bands. For a typical end-to-end simulation with $N_{\rm rec} = 6$, it is around 320 iterations.} The result of this procedure at the map level for the I and Q components, for a given realization, is shown in figures~\ref{fig:simu_galaxy_mapI} and~\ref{fig:simu_galaxy_mapQ}. The two figures display a sky reconstructed in 5 sub-bands within the 220~GHz wide band. The residual map is the difference between a reconstruction and the noiseless reference. 

We can see that the Galactic dust is efficiently reconstructed in the 5~sub-bands as the residuals are compatible with pure noise. Note that the noise is not white but has spatial correlations due to the deconvolution with the multi-peak synthesized beam (see section 3.1.1 in Hamilton \textit{et al.}~\cite{2020.QUBIC.PAPER1}). We also note that the residuals are higher on the edges than in the center of the sky patch. This is due to the higher coverage of the sky in the center due to the scanning strategy.

Instead of looking at the full map, the reconstructed intensity as a function of frequency can also be studied pixel by pixel.
\begin{figure}[t]
    \centering
    \includegraphics[width=0.9\linewidth]{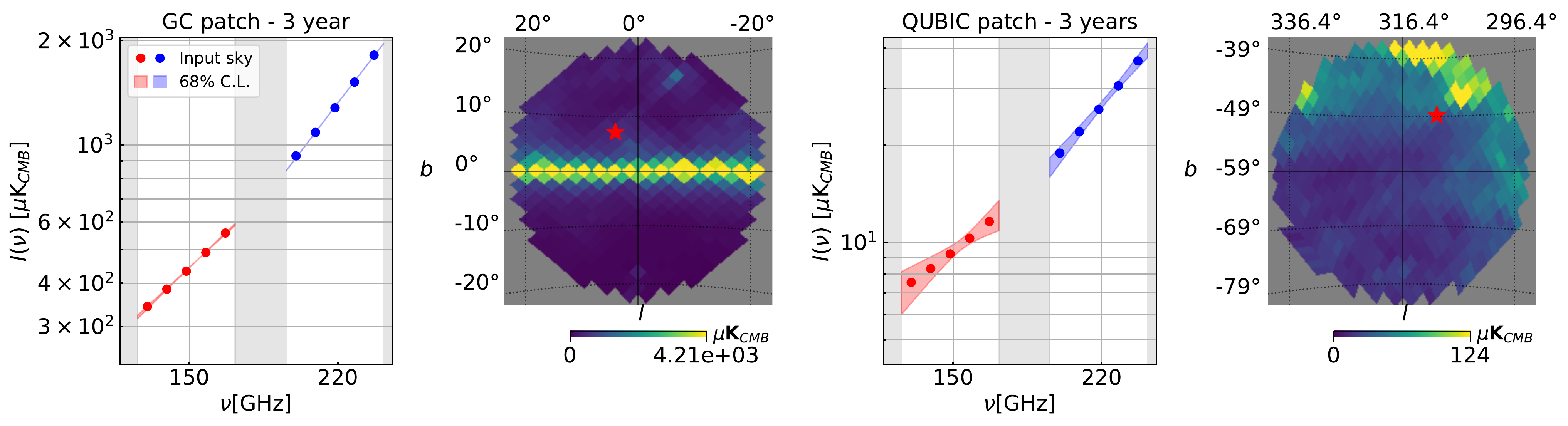} 
    \caption{Intensity as a function of the frequency for $N_{\rm rec}=5$~sub-bands in each wide band at 150~(red) and 220~(blue)~GHz for a given pixel. The grey regions correspond to the unobserved frequencies outside our physical bands. Two sky pixels are shown as red stars, one in a patch centered at the Galactic center and one in the patch that QUBIC plans to observe centered in [0, -57~deg]. Red and blue dots: Input sky convolved with the instrument beam. In both cases are shown in light color the 68\%~CL regions for a modified black-body spectrum reconstructed with a MCMC from our simulated measurements and sub-band covariance matrices (see figure~\ref{fig:bigcorrmatrix_maplevel} for the case of 3~sub-bands). Maps are in $\mu$K CMB and $N_{\rm side}=32$.}
    \label{fig:pix-pix_plot}
\end{figure}
Figure~\ref{fig:pix-pix_plot} shows the intensity of the input sky convolved with the instrument beam, and the reconstructed intensity for a given pixel, considering 5~sub-bands in each wide band at 150~(red) and 220~(blue)~GHz. We do not display the measured points and error-bars which are not good indicators of our uncertainties due to the highly anti-correlated nature of the covariance matrix (see figure~\ref{fig:bigcorrmatrix_maplevel} in the case of 3~sub-bands). Instead, we have performed a Monte-Carlo-Markov-Chain (MCMC) exploration of the amplitude and spectral-index of a typical dust model (modified black-body, see Irfan \textit{et al.}~\cite{irfan2019determining}) accounting for the sub-bands covariance matrix. The fit is done separately for our two physical bands at 150 and 220~GHz. The 68\%~confidence level (CL) is shown in light colors in each case and represents the QUBIC measurements within this band using spectral imaging. Note that the angular resolution of the maps improves with frequency (see section~\ref{ss:angular_reso}) and is not accounted for. As a result this simple analysis cannot be interpreted as a measurement of the dust spectral index which would require a more detailed analysis including the beam profile to properly infer the dust property in each sky pixel (such as~\cite{irfan2019determining}). This study is being carried out for a future article.

\begin{figure}[t]
    \centering
    \includegraphics[width=0.48\linewidth]{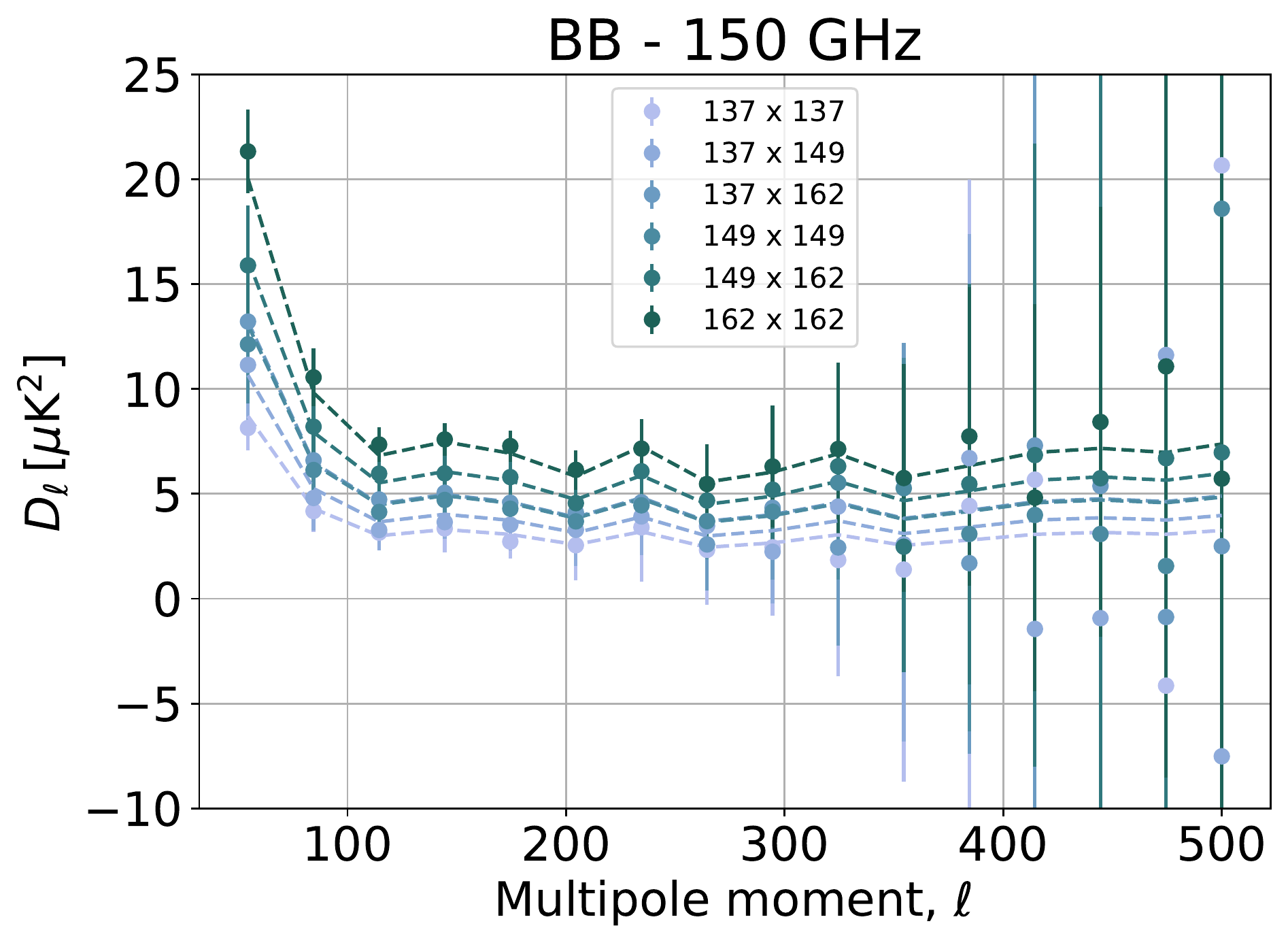}
    \includegraphics[width=0.48\linewidth]{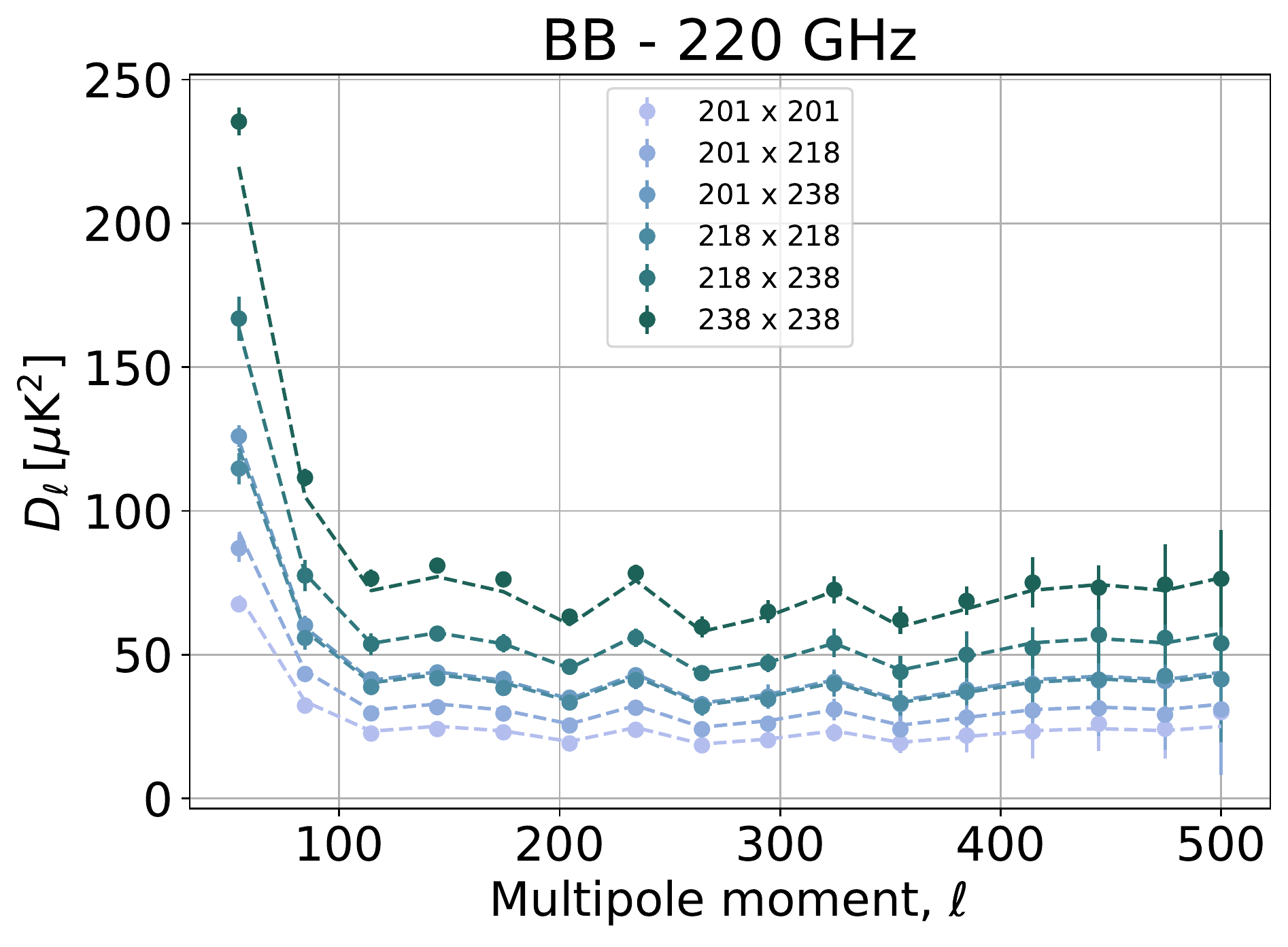}
    
    \caption{BB IBCS at 150~GHz (left) and 220~GHz (right) computed from reconstructed maps in 3~sub-bands (137, 149 and 162~GHz and 201, 218 and 238~GHz respectively), obtained with end-to-end simulations. For each IBCS, we cross-correlate 2~sub-bands with central frequencies in GHz shown in the legend of each plot. Dashed lines are the IBCS of the input sky that contains only Galactic dust. The dots with error bars show the mean and the standard deviation over 20 IBCS. Each IBCS is made with 2 maps with independent noise realizations to eliminate the noise bias.}
    \label{fig:spectra_3bands}
\end{figure}

The dust reconstruction is also studied in the angular power spectrum using the public code NaMaster~\cite{garcia2019disconnected} which computes TT, EE, BB and TE spectra where T is the temperature and E, B are the two polarization modes. Spectra are computed from a multipole moment $l=40$ to $l=2\times N_{\rm side}-1$ with $N_{\rm side} = 256$ the pixel resolution parameter for HEALPix maps. We compute inter-band cross spectra (IBCS), meaning that from $N_{\rm rec}$ sub-band maps, one can compute $N_{\rm rec}(N_{\rm rec}+1) / 2$ IBCS. Having independent noise realizations allows us to make IBCS crossing two realizations, so we eliminate the noise bias. BB IBCS for 3~sub-bands in each of the 150 and 220~GHz wide bands are shown in figure~\ref{fig:spectra_3bands}. We plot $D_l = \frac{l(l+1)}{2 \pi} C_l$, $C_l$ being the B-mode angular power spectrum. In this figure, the input theoretical dust spectra coming from the PySM model \texttt{d1} are superposed to the reconstructed ones.

\section{Noise characterization}\label{sec:noiseSection}
For the map-making described in section~\ref{ss:mapmaking}, we added noise to the TOD which was composed of the detector NEP ($4.7 \times 10^{-17}W/\sqrt{Hz}$) and photon NEP (see table~\ref{tab:parameters}). As before, the atmospheric load is assumed to be stable. The goal here is to study how close to optimal (in the statistical sense) is our spectral imaging map-making. We will study the noise behaviour as a function of the number of reconstructed sub-bands $N_{\rm rec}$. This is done at three different levels: on the reconstructed maps, on the power spectra computed from the maps, and on a likelihood estimation for the tensor-to-scalar ratio $r$. 

\subsection{Noise behaviour in the sub-bands at map level}\label{ss:noise_maplevel}

We performed simulations with 40 independent noise realizations and a noiseless simulation as a reference. After map-making, residual maps are computed by taking the difference between each simulation and the noiseless reference. 
For each pixel, we compute the covariance matrix, over all the noise realizations, between the sub-bands and the Stokes parameters. 
\begin{figure}[t]
    \centering
    \includegraphics[width= 0.48\linewidth]{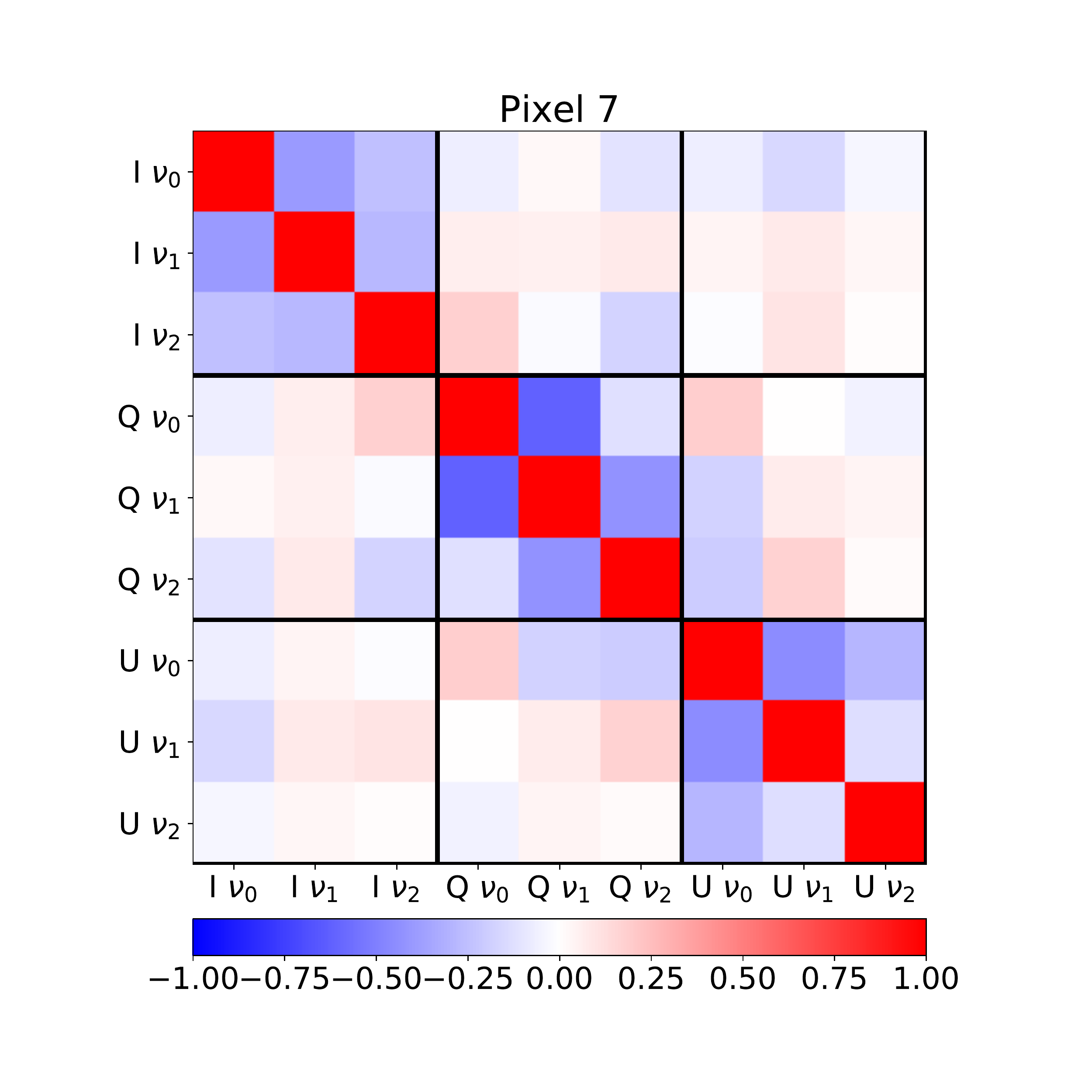}
    \includegraphics[width= 0.48\linewidth]{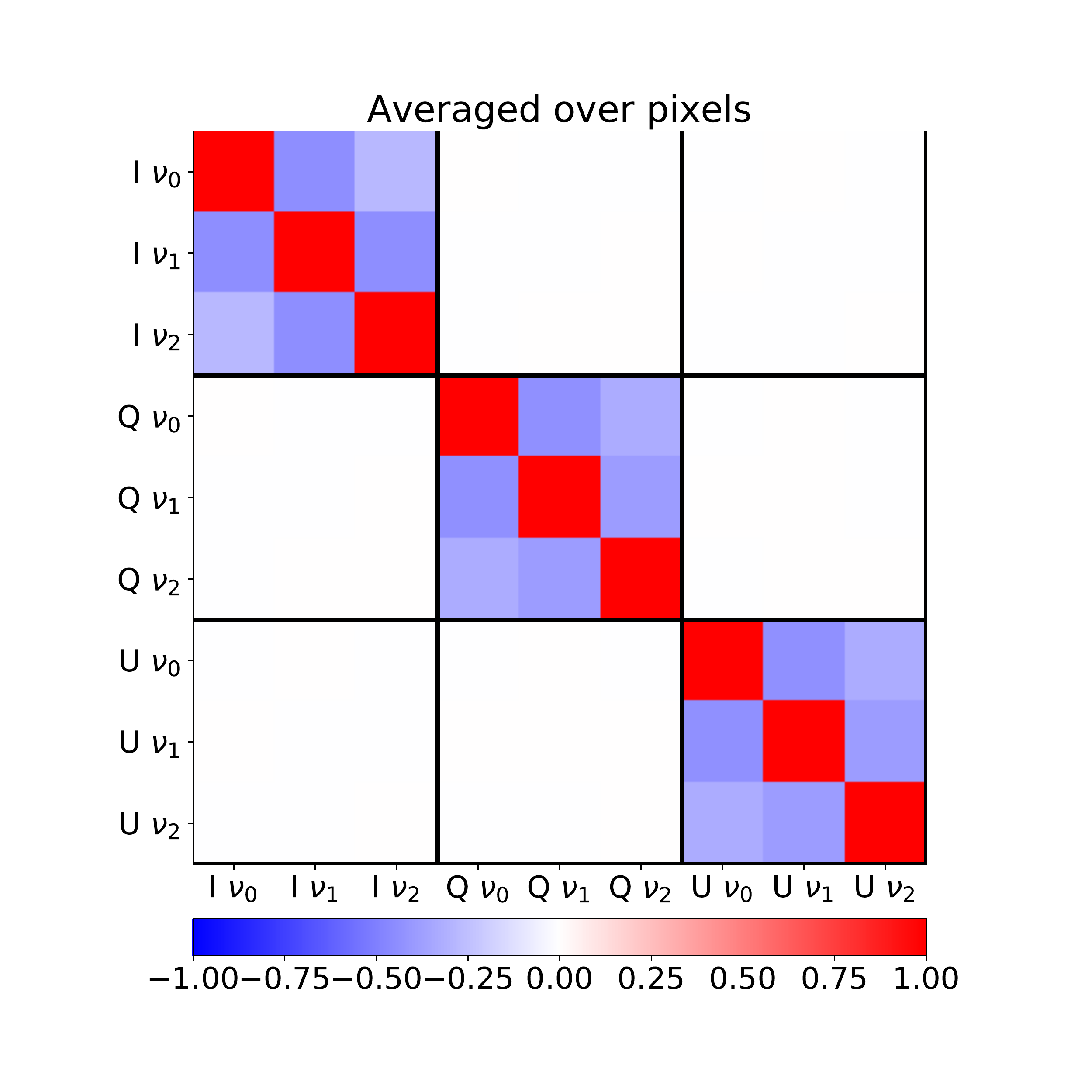}
    \caption{Correlation matrices between frequency sub-bands $\nu_0 = 137$~GHz, $\nu_1 = 149$~GHz and $\nu_2 = 162$~GHz and I, Q, U Stokes components obtained from 40 end-to-end simulations. \textit{Left:} example for a given pixel. \textit{Right:} The average over pixels. Blue means anti-correlations while red is for positive correlations.}
    \label{fig:bigcorrmatrix_maplevel}
\end{figure}

The reason why we treat the pixels separately, not computing covariance over them, is that the noise level varies with the position in the sky. This is due to the coverage of the sky by the instrument beam which is not uniform. Note that the QUBIC coverage is not trivial because of the multi-peak synthesized beam.

A correlation matrix for a given pixel, considering 3 sub-bands, is shown in figure~\ref{fig:bigcorrmatrix_maplevel} and we also show the average over pixels. It can be seen that for each Stokes parameter, residual sub-bands next to one another are anticorrelated and this is seen on every pixel. 
However, cross-correlations between Stokes parameters are negligible. This is why we can consider the 3 correlation matrices $C_{Ip}$, $C_{Qp}$ and $C_{Up}$ separately.

\subsection{Noise analysis using the power spectrum}
\label{ss:fastsim}
In section~\ref{ss:noise_maplevel} it was shown that a polychromatic interferometer has anti-correlations in neighbouring bands for each Stokes parameter. Furthermore, the spatial structure of noise is studied in detail in Hamilton \textit{et al.}~\cite{2020.QUBIC.PAPER1} using end-to-end simulations. This allows us to build a fast noise simulator that reproduces efficiently the noise behaviour in the reconstructed maps.
In the following, the fast noise simulator will be used in parallel with end-to-end simulations as it allows us to improve the statistics while consuming much less computing time.

\begin{figure}[t!]
    \centering
    \includegraphics[width=\linewidth]{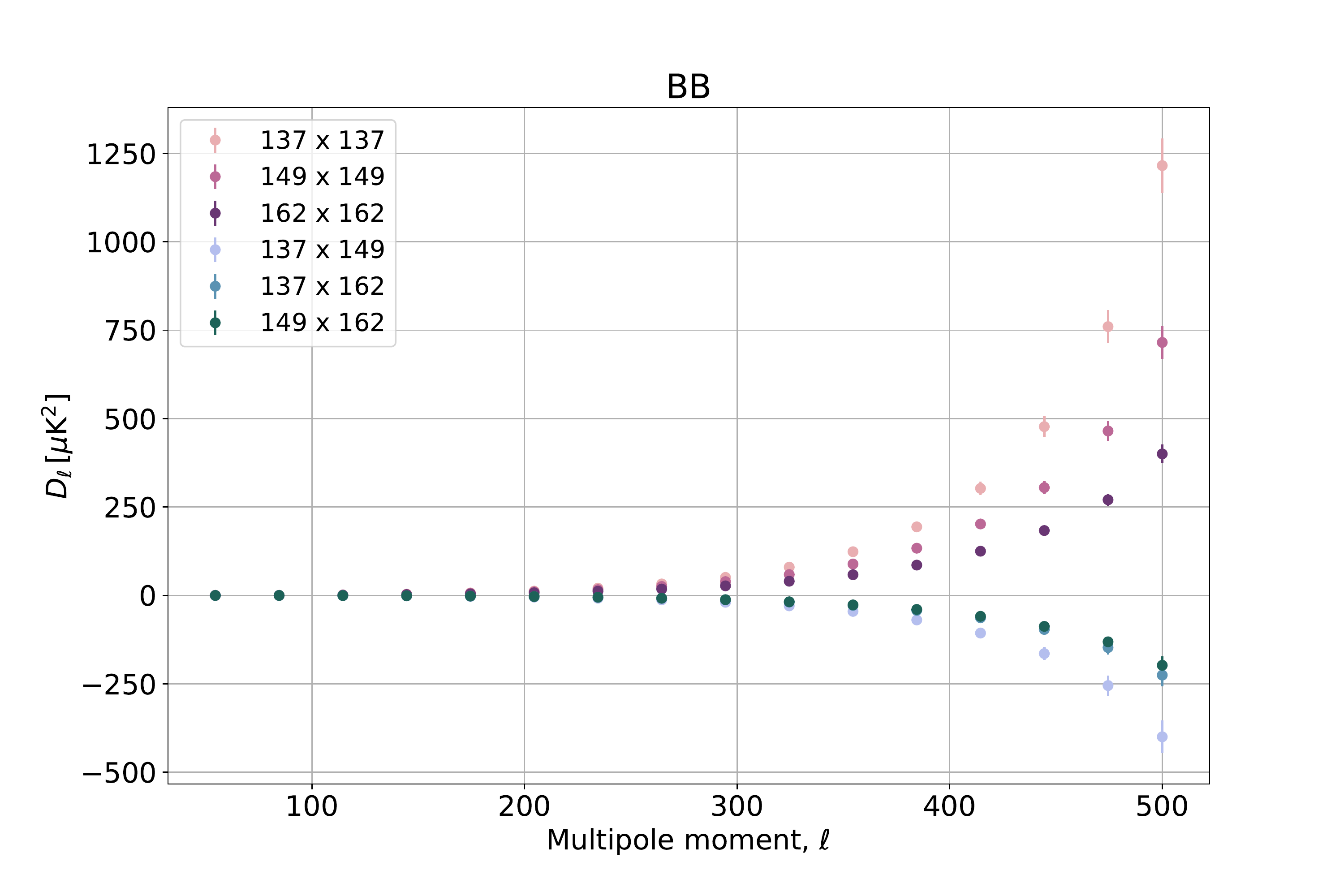}
    \caption{BB Inter Band Cross Spectra on the residual maps containing only noise for 3 sub-bands in the wide 150~GHz band, centered at 137, 149 and 162~GHz. Dots and error bars show average and standard deviation over 1000 independent noise realisation IBCS computed with the fast noise simulator.}
    \label{fig:IBCS_noise}
\end{figure}

We characterize the noise behaviour of spectral imaging using the power spectrum.
As shown in section~\ref{section:galaxy_dust}, from the maps we can compute power spectra using the public code NaMaster. From $N_{\rm rec}$ bands, we compute the IBCS for each TT, EE, BB and TE power spectra. As we are interested in the noise, we compute the power spectra of the residual maps containing only noise. Figure~\ref{fig:IBCS_noise} shows the IBCS computed for each noise realization in the case of 3~sub-bands at 150~GHz. As we plot $D_l$ the noise bias goes as $l(l+1)$. We find that the IBCS within the same band ($\nu_0\nu_0, \nu_1\nu_1$ and $\nu_2\nu_2$) are positively correlated. However the IBCS crossing 2~different bands ($\nu_0\nu_1, \nu_0\nu_2$ and $\nu_1\nu_2$) are anti-correlated.   

\begin{figure}[t!]
    \centering
    \includegraphics[width=0.95\linewidth]{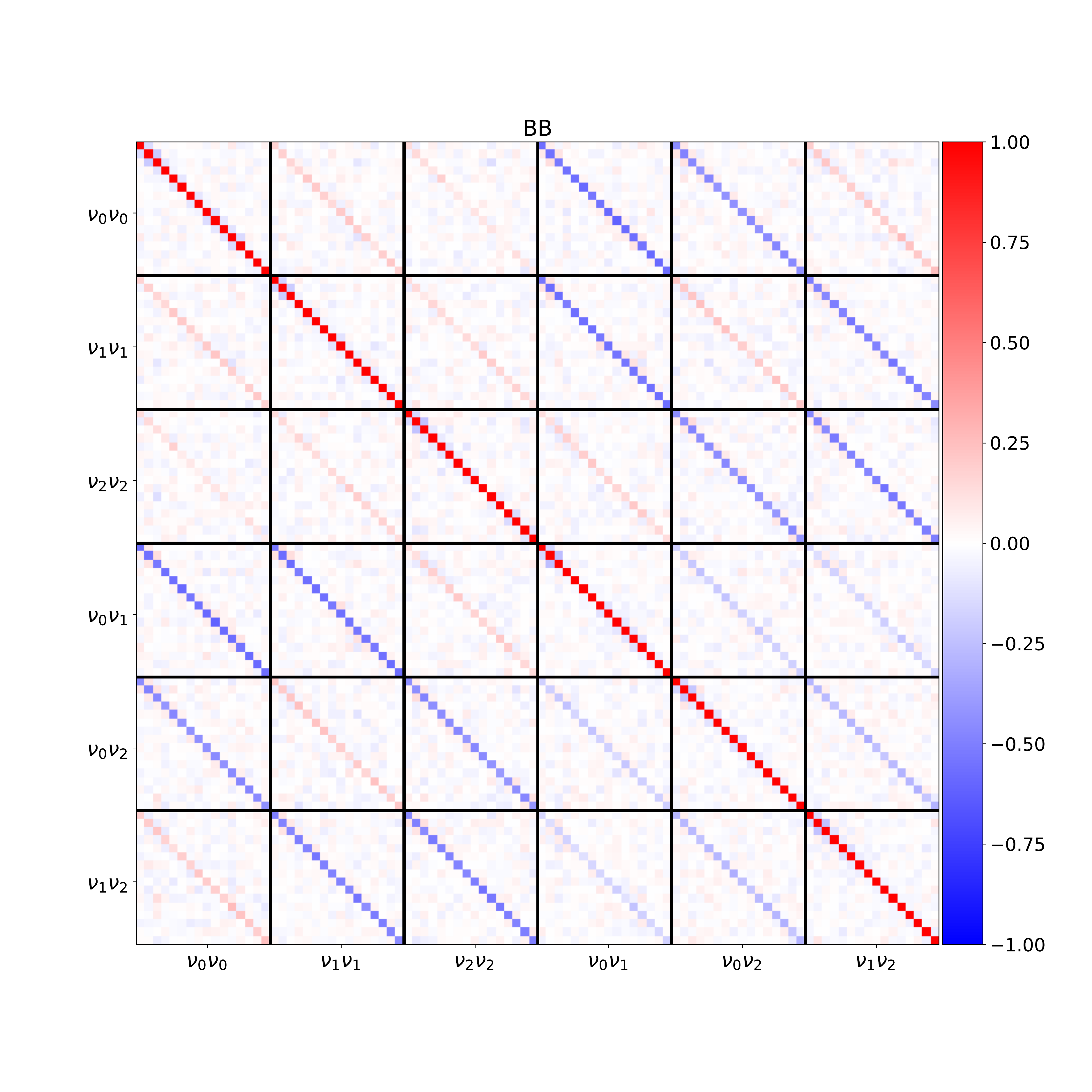}   
    \caption{Correlation matrix between $\ell$-bins and IBCS for BB angular power spectrum considering 3 sub-bands at $\nu_0=137$, $\nu_1=149$ and $\nu_2=162$~GHz. For example, $\nu_0\nu_1$ is the IBCS between frequencies $\nu_0$ and $\nu_1$. In each black square we show the correlations between the 16 $\ell$-bins used to compute the IBCS as in figure~\ref{fig:IBCS_noise}.}
    \label{fig:IBCS_BBcov}
\end{figure}

The correlations are observed in greater detail by computing the correlation matrices. In figure~\ref{fig:IBCS_BBcov}, we show the correlation matrix between $\ell$-bins and IBCS for BB angular power spectrum considering $N_{\rm rec}=3$~sub-bands in the 150~GHz wide band. In this matrix, we see that anti-correlations, in blue in the matrix, only appear between the IBCS crossing 2 different bands ($\nu_0\nu_1, \nu_0\nu_2$ and $\nu_1\nu_2$ in the case of 3~sub-bands) and that the correlations between bins are negligible. The same behaviour is observed for TT, EE and TE spectra. 

\subsection{Nearly optimal performance of spectral imaging}
\label{subsec:likelihood}
In order to assess, in a manner easy to interpret, how far from optimal is spectral imaging, we study how the tensor-to-scalar ratio $r$ is constrained as a function of the number of sub-bands~$N_{\rm rec}$. The sky model is a pure CMB sky, including lensing but no Galactic foregrounds, with $r=0$. The aim of this section is not to make precise forecasts on the sensitivity on $r$. This is addressed in the companion paper Hamilton \textit{et al.}~\cite{2020.QUBIC.PAPER1} and we expect to reach a sensitivity to B-modes corresponding to a 68\% CL upper-limit on the effective tensor-to-scalar ratio (primordial tensors + galactic dust) $\sigma(r) = 0.015$ with three years of observations.

The division of the wide band into a number of sub-bands for spectral imaging could have a detrimental effect on the estimate of the tensor-to-scalar ratio $r$ and this is what we want to quantify in this section. On the one hand, we would certainly like to make as many sub-bands as possible in order to constrain the foreground spectra in a very precise manner. However, on the other hand, there is an upper-limit to the achievable number of sub-bands, when the angular distance between peaks in the synthesized beams at different frequencies becomes smaller than the peak width (angular resolution), as explained in section~\ref{ss:SI-capabilities}. We therefore expect the performance of spectral imaging to degrade when projecting data onto too many sub-bands. In fact, even for a small number of sub-bands, spectral imaging cannot be strictly optimal because the synthesized beams at different sub-frequencies do not form an orthogonal basis. We therefore expect a certain loss in signal-to-noise ratio when performing spectral imaging. The higher the number of reconstructed sub-bands, the more overlap there is between the synthesized beam at each sub-frequency. This results in stronger degeneracy between sub-bands, hence a higher noise in the reconstruction. This is the price to pay for improved spectral resolution. As a result, one needs to find the best balance between performance and spectral resolution for a given scientific objective. 

\begin{figure}[t]
    \centering
    \includegraphics[width=\linewidth]{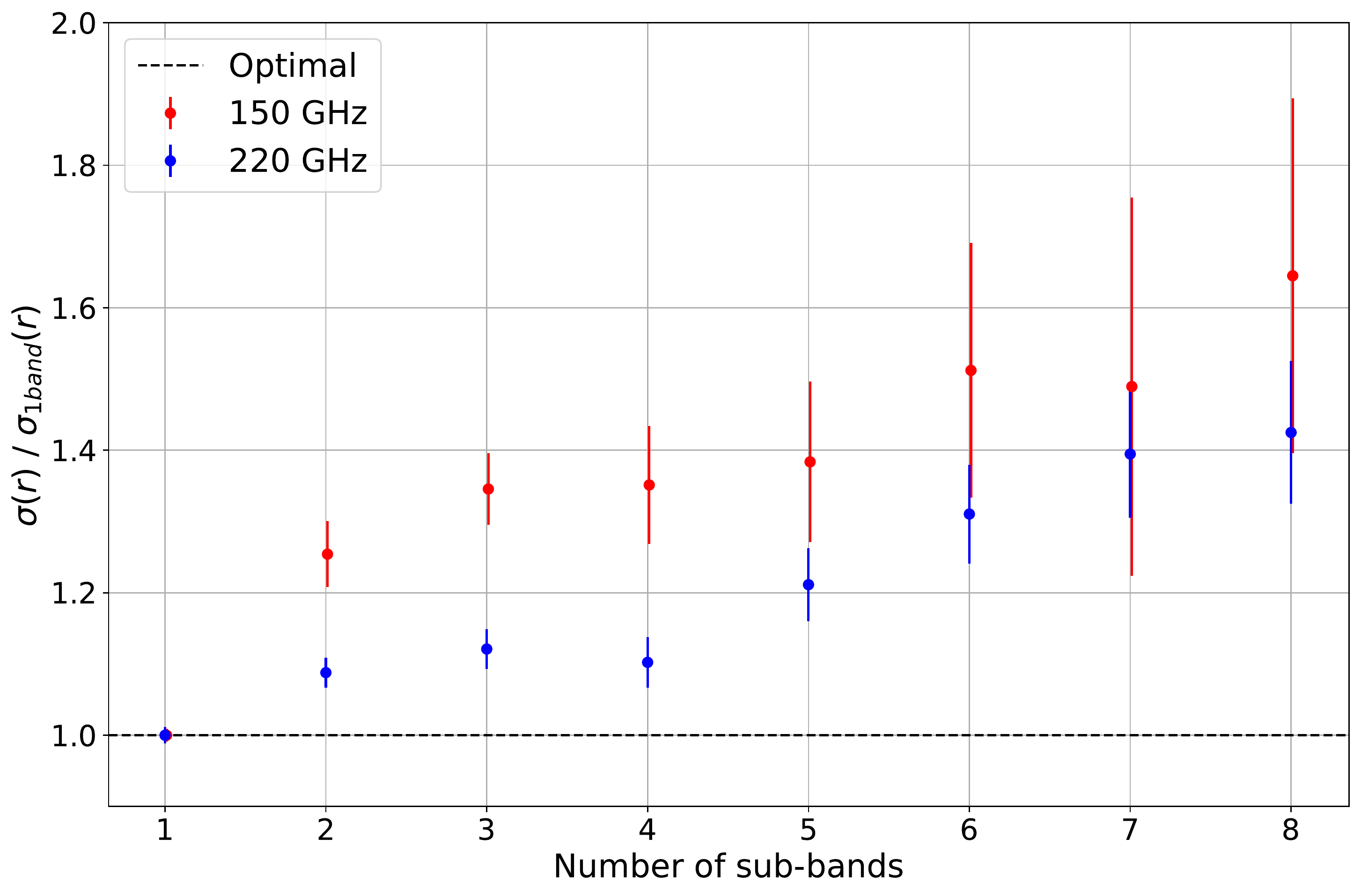}
     \caption{Uncertainties (68\% CL upper limits) on the tensor-to-scalar ratio $r$ obtained by combining an increasing number of sub-bands, normalized to that of one band, for a pure $r=0$ CMB (with lensing). The slow increase of the uncertainty on $r$ with the number of sub-bands illustrates the moderate sub-optimality of spectral imaging and shows that we can use up to 5 sub-bands with only 40\% degradation at 150~GHz (and only 20\% degradation at 220~GHz). It is possible to achieve 8 sub-bands but with more significant degradation.
    }
    \label{fig:likelihood_onlyCMB}
\end{figure}
For that purpose, we project simulated data onto an increasing number of sub-bands $N_{\rm rec}$, calculate the corresponding IBCS and in each case we compute a likelihood to estimate the tensor-to-scalar ratio $r$ combining all sub-bands accounting for their cross-correlations. From this method, we get the error on $r$ at 68\% confidence level for each number of sub-bands. This is presented in figure~\ref{fig:likelihood_onlyCMB}. We normalize by the case of ``spectral imaging'' with just one band. Error bars are obtained from a Monte-Carlo analysis, varying the data in the likelihood according to their diagonal uncertainties. As expected, we observe a moderate degradation due to spectral imaging in the sense that the constraints on $r$ become less stringent when the number of sub-bands is greater than one. This degradation slowly evolves from 25\% to 40\% at 150~GHz and from 10\% to 20\% at 220~GHz when the number of sub-bands evolves from 2 to 5. The better performance at 220~GHz is not a surprise as our horns are slightly multimoded at 220~GHz (see O'Sullivan \textit{et al.}~\cite{2020.QUBIC.PAPER8} for details) resulting in a flatter primary beam, which, in turn, favours spectral imaging because multiple peaks of the synthesized are higher in amplitude. It is possible to project onto as many as 8 sub-bands with a corresponding performance reduction due to the fact that synthesized beam peaks become too close with respect to their width, as explained in section~\ref{ss:SI-capabilities}.

This study demonstrates that, although not optimal from the noise point of view, spectral imaging performance remains close to optimal for up to 5 bands, providing extra spectral resolution that can be key for constraining foreground contamination with realistic models for which the spectrum might not be a simple power law, work being in progress on this. The appropriate balance between spectral resolution and noise performance can be adjusted for each specific analysis thanks to the fact that spectral imaging is done entirely in post-processing. 

This work is a first step to characterize the potential of spectral imaging. In order to place the technique in context, we need to include spectral imaging into a proper component separation step. This will be addressed in future work.

\section{Spectral imaging on real data}
\label{RealDatasection}

\begin{figure}[t!]
\centering
\includegraphics[width=0.9\hsize]{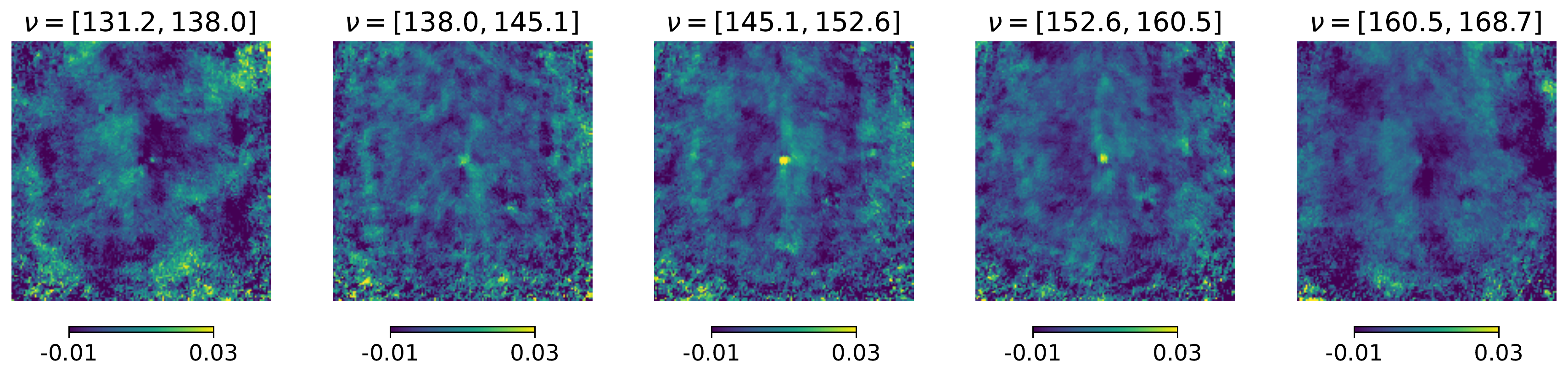}
   \caption{Calibration data with the source at 150 GHz projected on the sky using our map-making software to deconvolve from the multiple peaked synthesized beam and split the physical band of the instrument into 5 sub-bands.}
   \label{fig:real_data_maps}
\end{figure}

Spectral imaging has been applied on real data for the first time during the calibration campaign at the APC laboratory. The QUBIC instrument was placed on an alt-azimuth mount in order to scan a calibration source tuned at 150~GHz (with 144~Hz bandwidth) and placed in the far field. The corresponding analysis presented in the companion paper Torchinsky \textit{et al.}~\cite{2020.QUBIC.PAPER3}. We then perform a scan in azimuth and elevation with the instrument, obtaining a TOD for each bolometer. We then apply our spectral imaging map-making algorithm with five sub-bands to a selection of 26 bolometers that do not exhibit saturation. The synthesized beam for each bolometer is realistically modeled in our map-making through a series of Gaussian whose amplitude, width and locations are fit from a measured map of the synthesized beam for each bolometer (see figure 20 from Torchinsky \textit{et al.}~\cite{2020.QUBIC.PAPER3} for an example). The frequency evolution of this synthesized beam only assumes linear scaling with wavelength. We were able to reconstruct a map of the point-like artificial calibration source as well as its location in frequency space. In figure~\ref{fig:real_data_maps}, we show the reconstruction onto 5 sub-bands. The expected point-source shape is clearly visible in the central frequency sub-band containing the emission frequency of the source at 150~GHz.

\begin{figure}[ht!]
\centering
\includegraphics[width=0.8\hsize]{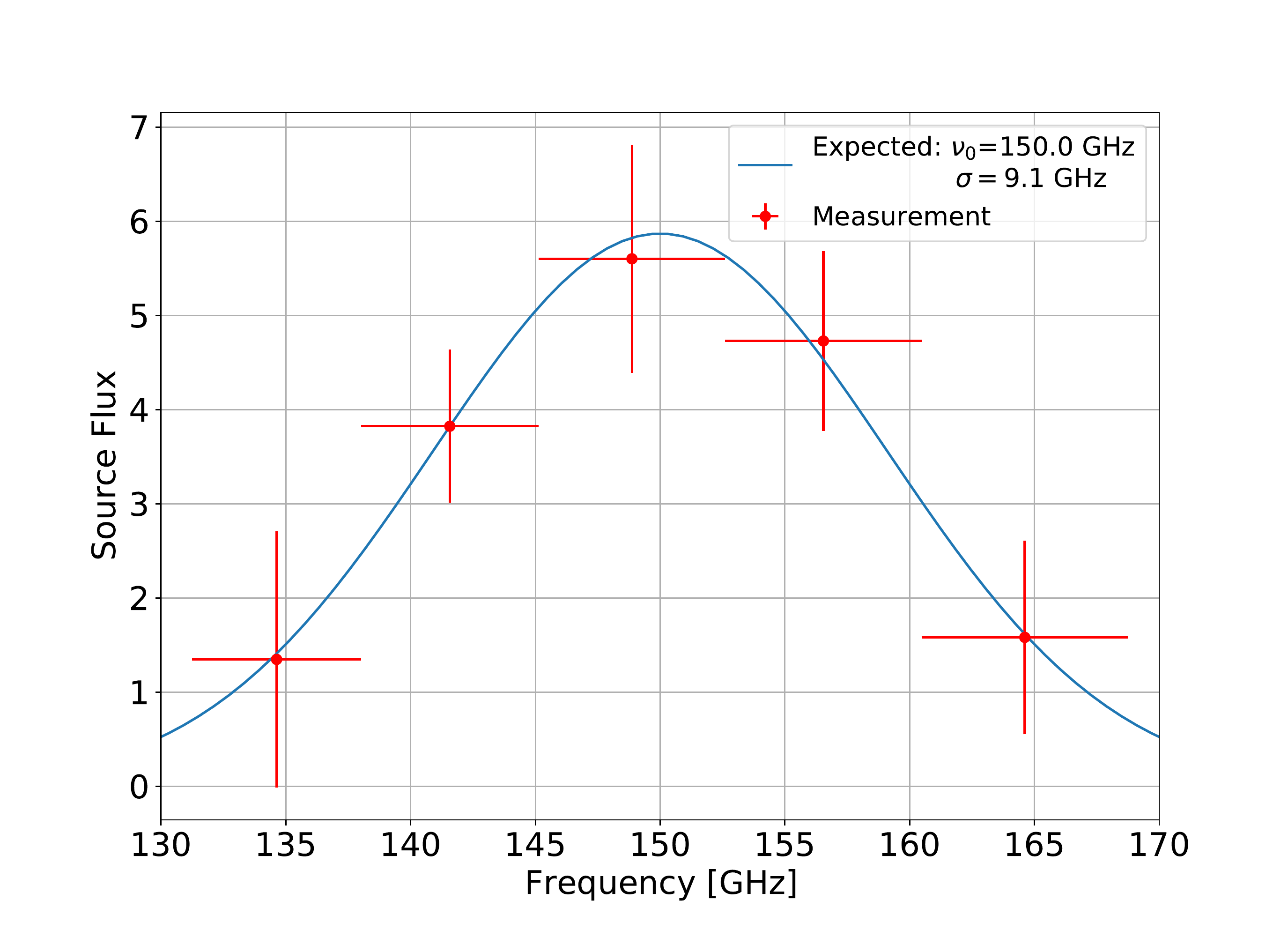}
   \caption{Measurement of the flux of the source in reconstructed sub-bands. The measurement (simple aperture photometry) in red is compared to the expected value spectrum in blue. The expected shape is a Gaussian centered on $\nu = 150$~GHz and a FWHM equal to $\frac{\nu}{(P-1)}$ with $P=8$ for the TD instrument. A global offset and the amplitude of the Gaussian are adjusted to the data.}
   \label{fig:real_data_reso}
\end{figure}

The calibration source is fainter in adjacent bands, and not visible in the furthest bands. In figure~\ref{fig:real_data_reso}, we show the detected amplitude in the central pixel as a function of the frequency. The measurement in red is compared to the expected value spectrum in blue. The expected shape is a Gaussian centered on $\nu = 150$~GHz and a FWHM equaled to $\frac{\nu}{(P-1)}$ as explained in section~\ref{ss:SI-capabilities}. Those data were acquired with the TD instrument that has a square $8\times 8$ feedhorn array so $P=8$. The global offset and the amplitude of the Gaussian are adjusted to the data. Error bars are computed in a very conservative manner.
 
\section{Conclusion}

In this article, we have shown how the new technique of bolometric interferometry offers the possibility to also perform spectral imaging. This makes it possible to split, in post processing, the wide-band observations into multiple sub-bands achieving spectral resolution. To illustrate this method, we apply it to the case of the QUBIC instrument soon to be installed at its observation site in Argentina.

After having presented the design, concept, and mathematical aspects of the instrument and the spectral imaging technique we have illustrated it on simple cases: monochromatic point sources, spatially extended sources, and sky maps with frequency-dependent emission such as Galactic dust. We have shown our ability to have increased spectral resolution with respect to the physical bandwidth, considering a full sky patch but also at the level of individual pixels. We studied the signal and noise behavior using Monte-Carlo simulations for an instrument like QUBIC which shows spatial and spectral correlations. We have quantified the loss of statistical performance for the measurement of the tensor-to-scalar ratio when increasing the number of sub-bands and have shown it to be moderate up to 5~sub-bands.

The precise measurement of foreground contaminants is essential for the detection of primordial B-modes.  Foregrounds have spectral properties distinct from the CMB which leads to the conclusion that only a multichroic approach enables the measurement and subtraction of foreground contamination. This is usually done in classical imagers through detectors operating at distinct frequencies, each of them being wide-band in order to maximize signal-to-noise ratio. However, constraining foregrounds with such data relies on extrapolation between distant frequency bands, which may miss non-trivial variations of the spectral behaviour of complex foregrounds such as multiple dust clouds in the line of sight. In particular, scenarios where dust exhibits a certain level of decorrelation between widely separated bands, or with non constant spectral indices would be impossible to be identified with a usual wide-band analysis. Spectral imaging could put significant constraints on such scenarios. This is being studied in detail by the QUBIC collaboration and will be presented in the near future.
 
In summary, spectral imaging improves spectral resolution within a wide physical band, while nearly preserving the optimal performance of the analysis. It may therefore become a key technique for detecting the elusive B-mode polarization of the CMB.

\section*{Acknowledgements}
We thank the anonymous referees as well as the editor for insightful reviews that greatly improved this work. This research used resources of the National Energy Research Scientific Computing Center (NERSC), a U.S. Department of Energy Office of Science User Facility located at Lawrence Berkeley National Laboratory, operated under Contract No. DE-AC02-05CH11231. QUBIC is funded by the following agencies. France: ANR (Agence Nationale de la Recherche) 2012 and 2014, DIM-ACAV (Domaine d’Interet Majeur-Astronomie et Conditions d’Apparition de la Vie), Labex UnivEarthS (Université de Paris), CNRS/IN2P3 (Centre National de la Recherche Scientifique/Institut National de Physique Nucléaire et de Physique des Particules), CNRS/INSU (Centre National de la Recherche Scientifique/Institut National des Sciences de l’Univers). Italy: CNR/PNRA (Consiglio Nazionale delle Ricerche/ Programma Nazionale Ricerche in Antartide) until 2016, INFN (Istituto Nazionale di Fisica Nucleare) since 2017. Argentina: MINCyT (Ministerio de Ciencia, Tecnología e Innovación), CNEA (Comisión Nacional de Energía Atómica), CONICET (Consejo Nacional de Investigaciones Científicas y Técnicas). 

\bibliographystyle{ieeetr}
\bibliography{qubic} 

\end{document}